\documentclass[lettersize,journal]{IEEEtran}
\IEEEoverridecommandlockouts
\usepackage{cite}
\usepackage{amsmath,amssymb,amsfonts}
\usepackage{graphicx}
\usepackage{textcomp}
\usepackage{xcolor}
\usepackage{algorithm}
\usepackage{algpseudocode}

\usepackage{booktabs,array}
\usepackage{subfigure}
\usepackage{bm}
\usepackage{comment}
\usepackage{color}
\usepackage{stfloats}
\usepackage{bm}
\usepackage{makecell}
\begin{document}
	
	\title{Movable IRS-Aided ISAC Systems: Joint Beamforming and Position Optimization\\
	}
	
	\author{\IEEEauthorblockN{Yue Geng, Tee Hiang Cheng, Kai Zhong \IEEEmembership{Graduate Student Member, IEEE}, \\ Kah Chan Teh, \IEEEmembership{Senior Member, IEEE}, and  Qingqing Wu, \IEEEmembership{Senior Member, IEEE}}
		\thanks{
			
			The work of Yue Geng, Tee Hiang Cheng, and Kah Chan Teh was supported by the School of Electrical and Electronic Engineering, Nanyang Technological University, Singapore. The work of Kai Zhong was supported by the National Natural Science Foundation of China under Grant 62501112 and in part by the China Postdoctoral Science Foundation under Grant 2025M773511. The work of Qingqing Wu was supported by NSFC under Grant 62371289 and Grant 62331022. \emph{Corresponding author: Kah Chan Teh; Kai Zhong}.
			
			Yue Geng, Tee Hiang Cheng, Kah Chan Teh are with the School of Electrical and Electronic Engineering, Nanyang Technological University, Singapore (Email: yue014@e.ntu.edu.sg; ethcheng@ntu.edu.sg; ekcteh@ntu.edu.sg).
			
			Kai Zhong is with the School of Information and Communication Engineering, University of Electronic Science and Technology of China, Chengdu, 611731, China (Email: 201921011206@std.uestc.edu.cn).
			
			Qingqing Wu is with the Department of Electronic Engineering, Shanghai Jiao Tong University, 200240, China (e-mail: qingqingwu@sjtu.edu.cn).
			
			This paper has been accepted for publication in IEEE Transactions on Wireless Communications (DOI: 10.1109/TWC.2026.3692892). 
			
			© 2026 IEEE. Personal use of this material is permitted. Permission from IEEE must be obtained for all other uses, in any current or future media, including reprinting/republishing this material for advertising or promotional purposes, creating new collective works, for resale or redistribution to servers or lists, or reuse of any copyrighted component of this work in other works.
			
		}

	}
	\maketitle
	
	\begin{abstract}
		Driven by intelligent reflecting surface (IRS) and movable antenna (MA) technologies, movable IRS (MIRS) has been proposed to improve the adaptability and performance of conventional IRS, enabling flexible adjustment of the IRS reflecting element positions. This paper investigates MIRS-aided integrated sensing and communication (ISAC) systems. The objective is to minimize the power required for satisfying the quality-of-service (QoS) of sensing and communication by jointly optimizing the MIRS element positions, IRS reflection coefficients, transmit beamforming, and receive filters. To balance the performance-cost trade-off, we proposed two MIRS schemes: element-wise control and array-wise control, where the positions of individual reflecting elements and arrays consisting of multiple elements are controllable, respectively. To address the joint beamforming and position optimization, a product Riemannian manifold optimization (PRMO) method is proposed, where the variables are updated over a constructed product Riemannian manifold space (PRMS) in parallel via penalty-based transformation and Riemannian Broyden–Fletcher–Goldfarb–Shanno (RBFGS) algorithm. Simulation results demonstrate that the proposed MIRS outperforms conventional IRS in power minimization with both element-wise control and array-wise control. Specifically, with different system parameters, the minimum power is achieved by the MIRS with the element-wise control scheme, while suboptimal solution and higher computational efficiency are achieved by the MIRS with array-wise control scheme.
	\end{abstract}
	
	
	\section{Introduction}
	Integrated sensing and communication (ISAC) has emerged as a key enabling technology for sixth-generation (6G) wireless networks, aiming to unify radar sensing and communication functions within a shared platform \cite{6gisac}. By leveraging common spectrum, hardware, and signal processing frameworks, ISAC supports high-speed data transmission while simultaneously providing high-accuracy sensing capabilities \cite{ISAC1,ISAC2}. This joint design not only improves spectral and energy efficiency but also reduces system cost and complexity through resource reuse. As the integration inherently introduces coupling between the two functions, extensive works have been implemented to carefully balance performance trade-offs via advanced waveform design, beamforming, and interference management strategies \cite{ISAC3}.  
	
	The performance of ISAC is fundamentally limited by complex wireless channels, particularly blockage effects that severe line-of-sight (LoS) links. To tackle this problem, intelligent reflecting surface (IRS) has emerged as a promising technology for enhancing ISAC systems \cite{IRSISAC}. By carefully designing the reflection coefficients of the passive reflecting elements within IRS, the wireless environment of ISAC systems could be improved, leading to enhanced communication throughput and sensing performance \cite{IRS1,IRS2,IRS3}. However, a large number of reflecting elements may be required for traditional IRS equipped with elements arranged in fixed-position array (FPA), which poses challenges in terms of the physical device size and the precise control of the reflection coefficients for each element. Besides, such FPA-enabled IRS (FPA-IRS) lacks the capability to adjust the incident and reflected channels. Since IRS-assisted links suffer from double fading, the performance gains for wireless communication systems brought by IRS may be limited. To directly optimize the wireless channel, movable antenna (MA) technology has been proposed recently. Unlike conventional FPA, whose configurations remain unchanged after deployment, MAs can dynamically reposition their elements within a confined region spanning several wavelengths, enabling real-time adaptation to varying channel conditions and task requirements \cite{MA1,MA2,MA3,MA4}. Recent works have implemented MA in ISAC and IRS-aided ISAC systems, which demonstrate the advantages of MA over FPA in enhancing ISAC performance \cite{MAISAC1,MAISAC2,MAIRSISAC1,MAIRSISAC2}. However, MA is typically equipped with the base station (BS) in existing works. Although such implementation can improve the BS-side channel, it cannot inherently alter the propagation environment at the reflecting point, leaving the potential of spatial diversity largely unexploited.
	
	To overcome the above disadvantages of conventional FPA-IRS, the concept of movable IRS (MIRS) has recently attracted significant attention. Compared with conventional IRS with fixed reflecting elements, MIRS introduces an additional spatial degree of freedom (DoFs) by allowing the reflecting elements to move within a confined wavelength-scale region. By jointly exploiting phase control and position adjustment, MIRS can further improve the reflected channel conditions and enhance the effective channel gain. Similar to conventional IRS, MIRS has a wavelength-scale size and can be deployed in indoor environments, on building surfaces, or on autonomous aerial vehicles-mounted platforms. A movable intelligent surface (MIS) architecture was proposed to enable dynamic beamforming for enhancing communication systems in \cite{MIS}, where a smaller movable metasurface is globally shifted relative to a fixed metasurface to synthesize diverse beam patterns through superimposed phase variations, eliminating the need for element-wise phase control. By further exploiting the MA architecture and the phase-control capability of IRS, existing studies have proposed the MIRS architecture, in which the IRS reflecting elements are mounted on movable modules, enabling the joint control of both the positions of the reflecting elements and their reflection coefficients. In \cite{MEIRS}, a one-dimensional MIRS with movable elements was proposed to tackle the phase distribution offset problem in traditional FPA-IRS-aided communication systems. The joint beamforming and position optimization for 2-dimensional (2D) MIRS-aided communications was studied in \cite{MEIRS2}, where the positions and reflection coefficients of the IRS elements are jointly controlled to improve the communication rate for a single user, leading to superior performance compared to FPA-IRS. Similar architecture was implemented in \cite{MEIRS3}, where the MIRS was utilized for enhancing the throughput of multi-user communications, and it has been shown to provide superior performance as compared with FPA-IRS. For ISAC systems, an MIRS-aided ISAC system was studied in \cite{MEIRSISAC}, where the sensing signal-to-clutter-and-noise-ratio for one target was maximized with the constraints of communication rates lowerbound, and the positions of IRS elements were optimized via a memory-penalized projected gradient descent (MPPGD) algorithm.
	
	Although existing works have addressed several challenges in MIRS-aided ISAC systems, a number of critical issues remain unresolved. First, future ISAC systems are expected to simultaneously support multiple communication users and multiple sensing targets, such as in vehicular networks and internet of things systems \cite{MT1, MT2, MT3}. While IRS has been widely investigated to enhance ISAC performance by creating controllable reflected channels, MIRS introduces additional spatial DoFs that can further improve the reflected channel conditions. However, the potential of MIRS in multi-user multi-target ISAC systems has not yet been fully investigated. Second, in radar sensing tasks such as target detection and target tracking, the sensing signal-to-interference-plus-noise ratio (SINR) is a key performance metric since it directly determines detection reliability and estimation accuracy \cite{SINR1,SINR2}. However, although recent studies have explored MIRS-aided ISAC systems, they typically adopt alternative sensing metrics such as beam pattern gain \cite{MEIRSISAC}. The SINR-based sensing performance in MIRS-aided multi-user multi-target ISAC systems has not yet been systematically investigated. To address the above issues, we investigate an MIRS-aided multi-user multi-target ISAC system in this work, and we aim to minimize the total transmit power required for the system while satisfying the communication rate requirements of communication users (CUs) and the sensing SINR constraints of multiple targets. To solve this problem, we propose a product Riemannian manifold optimization (PRMO) method to obtain feasible solutions by jointly and simultaneously optimizing all variables, thereby achieving more favorable solutions than that obtained by optimizing the variables separately. In addition, existing studies on MIRS mainly focus on position optimization for individual IRS elements, which incurs significant computational overhead when precisely optimizing the position of each element. Besides such element-wise control, we investigate an array-wise control scheme for MIRS, wherein arrays composed of multiple reflecting elements are formed, and the positions of the arrays, rather than individual elements, are controlled to reduce the computational overhead. The main contributions are summarized as follows:
	\begin{itemize}
		\item We establish the system model and investigate the power minimization problem for an MIRS-aided ISAC system, aiming to minimize the power required for satisfying the communication rate and sensing SINR thresholds for multiple CUs and targets. To the best of our knowledge, the problem of MIRS-aided multi-user multi-target ISAC with SINR-based sensing constraints has not yet been investigated. With both element-wise control and array-wise control modes for the MIRS, we proposed a PRMO method to obtain feasible solutions by jointly optimizing the beamforming, receive filters, IRS reflection coefficients, and position variables.
		\item Simulation results show that the proposed MIRS under both control modes can achieve superior performance compared with FPA-IRS. Although the MIRS with element-wise control achieves the lowest transmit power under the same constraints, MIRS with array-wise control achieves sub-optimal performance while reducing computational time, thus offering an approach for balancing performance and computational cost. The results also verify the effectiveness of the proposed method under imperfect channel state information (CSI) condition.
	\end{itemize}
	
	The rest of the paper is organized as follows. Section II introduces the system model. The performance metrics and problem formulation are given in Section III. Section IV develops the PRMO for solving the problem. Section V provides simulation results. Section VI concludes the paper.
	
	\textit{Notations:}  Scalars, vectors, and matrices are indicated as $a$, $\mathbf{a}$, and $\mathbf{A}$, respectively. $\mathbf{a}[n]$ denotes the $n$-th element of $\mathbf{a}$. $\mathbf{A}[n,m]$ is the element in the $n$-th row and $m$-th column of $\mathbf{A}$, $\mathbf{A}_{:n}$ and $\mathbf{A}_{n:}$ denote the $n$-th column and $n$-th row of $\mathbf{A}$, respectively. $\mathbf{A}^T$, $\mathbf{A}^H$ and $\mathbf{A}^*$ indicate the transpose, conjugate transpose, and conjugate of $\mathbf{A}$, respectively. $\Re(\cdot)$ denotes taking the real part. $\mathbf{I}$ indicates identity matrix. $\odot$ denotes the element-wise multiplication. $\lvert \cdot \rvert$, $\lVert \cdot \rVert_2$, and $\lVert \cdot \rVert_F$ indicate modulus, 2-norm and Frobenius norm, respectively. $\operatorname{diag}(\mathbf{a})$ and $\operatorname{diag}(\mathbf{A})$ represent the diagonalization of $\mathbf{a}$ and the main diagonal of $\mathbf{A}$, respectively. $\operatorname{ddiag}(\mathbf{A})$ is the diagonal matrix formed from the diagonal elements of $\mathbf{A}$. $\operatorname{vec}(\cdot)$ and $\operatorname{Tr}(\cdot)$ are the vectorization and trace of a matrix, respectively.

	\section{System Model}
	In this section, we introduce the system model of the MIRS-aided multi-user multi-target ISAC system. As shown in Fig. 1, a BS equipped with $M$ transmit and receive antennas simultaneously performs radar sensing for $K_t$ targets and serves $K_c$ single-antenna CUs for communications. The antennas are arranged as a uniform linear array (ULA) with half-wavelength spacing, and the positions of the antennas with respect to the reference position $b_0=0$ are given as $\mathbf{b} = [b_1,\dots,b_M]^T\in\mathbb{R}^M$, where $b_m$ denotes the position of the $m$-th antenna. Since the direct BS-UE channel is blocked, an MIRS with $N$ movable reflecting elements confined within a 2D region is deployed to provide cascaded channel links. Denote the sets of CUs and targets as $\mathcal{K}_c\triangleq\{1,\dots,K_c\}$ and $\mathcal{K}_t\triangleq\{1,\dots,K_t\}$, respectively. In the following, we investigate the channel modeling for the MIRS-aided ISAC systems. Specifically, two control modes for the MIRS including the element-wise control and the array-wise control are considered.
	\begin{figure}[t]
		\centering{\includegraphics[width=1\columnwidth]{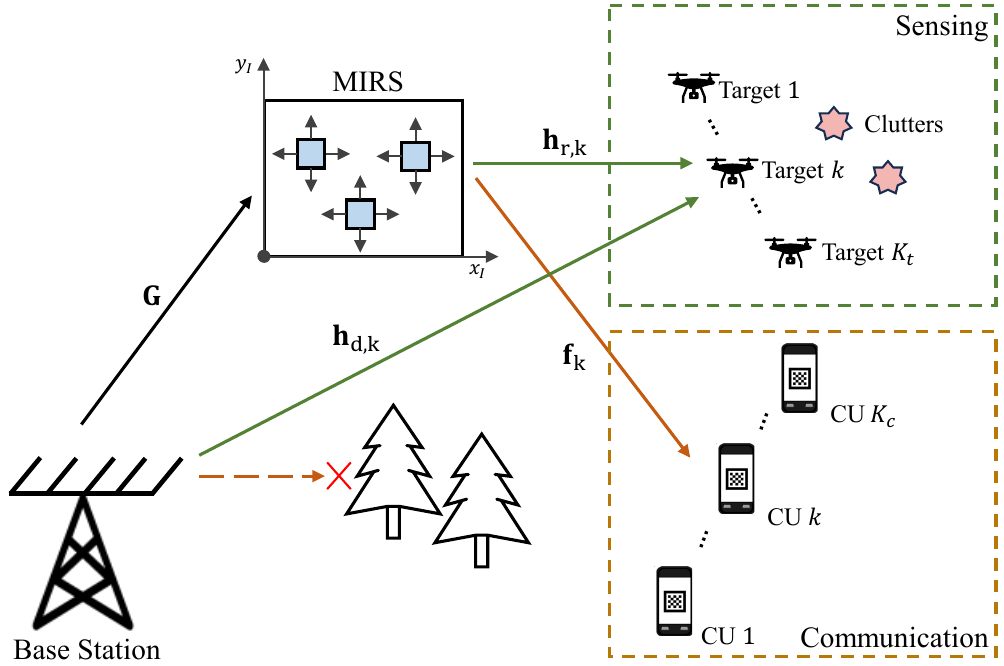}}
		\caption{The proposed MIRS-aided ISAC system.}
	\end{figure}
	\subsection{Channel Modeling for MIRS With Element-Wise Control}
	Firstly, we consider the element-wise control for the MIRS, where the position of each element is independently controlled. Since the MIRS architecture is more effective in multipath environments, we assume that multipath channel information is available in the system \cite{MA3}. Meanwhile, the distances between the BS, the MIRS, and the users are assumed to be sufficiently large to satisfy the far-field propagation condition. Under this condition, the angle of departure (AoD), angles of arrival (AoA), and complex response of each channel path are identical for different antennas in BS region and movable elements in MIRS region, while only the phases of the multi-path channels are different for the positions of BS antennas and MIRS elements. Based on the above, we propose the following channel modeling.
	\subsubsection{Communication Channels}
	Let $\mathbf{G}\in\mathbb{C}^{N\times M}$ and $\mathbf{f}_k\in\mathbb{C}^{N\times 1}$ denote the BS-IRS channel and the channel between IRS and the $k$-th CU, respectively. In this paper, we exploit the geometry model for modeling the channels, where each transmit path arrives at the receiver through only one receive path \cite{MA3,MA4}. We assume that $L$ transmit and receive paths exist for the BS-IRS and IRS-CU channels. Based on the above, the BS-IRS channel is given as
	\begin{equation}
		\label{G}
		\mathbf{G} = \mathbf{F}_a(\mathbf{t})^H\mathbf{\Sigma}_\mathrm{G}\mathbf{G}_t(\mathbf{b}),
	\end{equation}
	where the terms are defined as follows:
	\begin{itemize}
		\item $\mathbf{\Sigma}_\mathrm{G} =\operatorname{diag}([\sigma_{G,1},\dots,\sigma_{G,L}]^T)\in \mathbb{C}^{L \times L}$ denotes the path response matrix of the BS-IRS channel, where $L$ is the number of channel paths between the BS and MIRS, $\sigma_{G,l}$ denotes the complex response of the $l$-th path.
		\item $\mathbf{G}_t(\mathbf{b}) = \left[\mathbf{g}(b_1),\dots,\mathbf{g}(b_M)\right]\in\mathbb{C}^{L\times M}$ denotes the field response matrix (FRM) of the BS. As given in \cite{MA4}, the normalized wave vector of the $l$-th path can be defined as $\mathbf{n}_{t,l}=[\sin \theta_{t,l}\cos \phi_{t,l},\cos \theta_{t,l},\sin \theta_{t,l}\sin \phi_{t,l}]^T$,	where $\theta_{t,l}\in[0,\pi]$ and $\phi_{t,l}\in[0,\pi]$ are the elevation and azimuth AoDs of the $l$-th path. Denote the three-dimensional (3D) coordinate vector of the $m$-th antenna and the origin point of the transmit antenna array as $\mathbf{b}_m=[0,b_m,0]^T$ and $\mathbf{b}_o=[0,0,0]^T$, respectively. Then, the difference of the signal propagation for the $l$-th path between the position $\mathbf{b}_m$ and the reference point $\mathbf{b}_o$ is $\mathbf{n}_{B,l}^T\mathbf{b}_m = \cos \theta_{t,l} b_m$. Based on the above, the field response vector (FRV) of each transmit antenna is $\mathbf{g}({b}_m)=\left[e^{\jmath\frac{2\pi}{\lambda}{\rho}_{t,1}b_m},\dots,e^{\jmath\frac{2\pi}{\lambda}{\rho}_{t,L}b_m}\right]^T\in\mathbb{C}^{L},\forall m$, where ${\rho}_{t,l}=\cos \theta_{t,l}$. Then $\mathbf{G}_t(\mathbf{b})$ is obtained by stacking the FRV of the $M$ transmit antennas. $\mathbf{G}_t$ is a constant matrix since the BS antennas are fixed.
		\item $\mathbf{F}_a(\mathbf{t}) = \left[\mathbf{f}_a(\mathbf{t}_1),\dots,\mathbf{f}_a(\mathbf{t}_N)\right]\in\mathbb{C}^{L\times N}$ denotes the FRM of the MIRS for the arrival paths. Denote the 3D positions of the $n$-th element of the MIRS and the origin of the MIRS as $\bar{\mathbf{t}}_n=[t_{x,n},t_{y,n},0]^T$ and $\bar{\mathbf{t}}_0=[0,0,0]^T$. The normalized wave factor of the $l$-th receive path is $\mathbf{n}_{r,l}=[\sin \theta_{r,l}\cos \phi_{r,l},\cos \theta_{r,l},\sin \theta_{r,l}\sin \phi_{r,l}]^T$, then the signal propagation difference between the $n$-th movable element and the reference point of the MIRS is $\mathbf{n}_{r,l}^T\bar{\mathbf{t}}=\sin \theta_{r,l}\cos \phi_{r,l}t_{x,n}+\cos \theta_{r,l}t_{y,n}$. The FRV of the $n$-th element for the receive paths is $\mathbf{f}_a(\mathbf{t}_n)=\left[e^{\jmath\frac{2\pi}{\lambda}\boldsymbol{\rho}_{r,1}^T\mathbf{t}_n},\dots,e^{\jmath\frac{2\pi}{\lambda}\boldsymbol{\rho}_{r,L}^T\mathbf{t}_n}\right]^T\in\mathbb{C}^{L},\forall n$, where $\mathbf{t}_n=[t_{x,n},t_{y,n}]^T$, $\boldsymbol{\rho}_{r,l} = [\sin \theta_{r,l}\cos \phi_{r,l},\cos \theta_{r,l}]^T$, $\theta_{r,l}\in[0,\pi]$ and $\phi_{r,l}\in[0,\pi]$ are the elevation and azimuth AoAs of the $l$-th path, respectively. $\mathbf{F}_a(\mathbf{t})$ can be adjusted by controlling the element positions of the MIRS.
	\end{itemize}
	
	The set of element positions of the MIRS is given as $\mathbf{t}=[\mathbf{t}^T_1,\dots,\mathbf{t}^T_N]\in\mathbb{R}^{2N}$. The positions are confined in the 2D square region $\mathcal{C}$ with edge lengths of $A$. To avoid the coupling effect between the antennas, the element positions of the MIRS should satisfy $\lVert \mathbf{t}_n-\mathbf{t}_{n'} \rVert\geq \lambda/2$, where $\lambda$ is the signal wavelength. Based on the above, $\mathbf{G}$ is a function of $\mathbf{t}$, and each element of which can be calculated as
	\begin{equation}
		\mathbf{G}(\mathbf{t})[n,m] =\sum_{l=1}^{L} \sigma_{G,l}e^{\jmath\frac{2\pi}{\lambda}({\rho}_{t,l}{b}_m-\boldsymbol{\rho}_{r,l}^T\mathbf{t}_n)},\forall m, n.
	\end{equation}
	
	For simplicity, we assume that the AoD and AoA at each reflecting element are supplementary with respect to the surface normal. Then, the FRM of the MIRS for the reflecting paths is given by $\mathbf{F}_d(\mathbf{t}) = \left[\mathbf{f}_d(\mathbf{t}_1),\dots,\mathbf{f}_d(\mathbf{t}_N)\right]\in\mathbb{C}^{L\times N}$, where the FRV of each ME is calculated as $\mathbf{f}_d(\mathbf{t}_n)=\left[e^{-\jmath\frac{2\pi}{\lambda}\boldsymbol{\rho}_{r,1}^T\mathbf{t}_n},\dots,e^{-\jmath\frac{2\pi}{\lambda}\boldsymbol{\rho}_{r,L}^T\mathbf{t}_n}\right]^T\in\mathbb{C}^{L}, \forall n$. Similar to \eqref{G}, the channel between the MIRS and the $k$-th CU is given by
	\begin{equation}
		\mathbf{f}_k(\mathbf{t}) = \mathbf{F}_d(\mathbf{t})^H\mathbf{\Sigma}_{\mathrm{f},k}\mathbf{1},
	\end{equation}
	where the path response matrix of the IRS-CU channel is given by $\mathbf{\Sigma}_{\mathrm{f},k} =\operatorname{diag}([\sigma_{k,1},\dots,\sigma_{k,L}]^T) \in \mathbb{C}^{L \times L}, \forall k$ with complex response $\sigma_{k,l}$ for the $l$-th path. Based on the above, each element of $\mathbf{f}_k$ can be calculated as
	\begin{equation}
		\mathbf{f}_k(\mathbf{t})[n] =\sum_{l=1}^{L} \sigma_{k,l}e^{\jmath\frac{2\pi}{\lambda}(\boldsymbol{\rho}_{r,l}^T\mathbf{t}_n)},\forall n.
	\end{equation}
	
	Denote the phase shift matrix of the MIRS as $\mathbf{\Phi} = \operatorname{diag}(e^{j\phi_1},\dots,e^{j\phi_N})$, where $\phi_n \in [0,2\pi), \forall n$ is the phase shift of each element. To facilitate derivations, we define the phase shift vector of the MIRS as $\boldsymbol{\phi} = [e^{j\phi_1},\dots, e^{j\phi_N}]^H\in\mathbb{C}^N$, then the equivalent channel between the BS and the $k$-th CU can be denoted as
	\begin{equation}
		\label{Ch1}
		\mathbf{h}^H_{C,k}(\mathbf{t},\boldsymbol{\phi}) \!=\! \mathbf{f}^H_{k}(\mathbf{t})\mathbf{\Phi}\mathbf{G}(\mathbf{t})\!=\!\boldsymbol{\phi}^H\operatorname{diag}(\mathbf{f}^H_{k}(\mathbf{t}))\mathbf{G}(\mathbf{t}), \forall k \!\in\! \mathcal{K}_c,
	\end{equation}
	which is controllable by adjusting $\mathbf{t}$ and $\boldsymbol{\phi}$.

	\subsubsection{Sensing Channels}
	For the radar sensing, we assume the point target case, where the target is regarded as a single scatter with a small spatial extent. As commonly adopted in radar sensing systems, the channels between the BS/MIRS and the targets are assumed to be LoS. Note that the geometric channel is equivalent to the LoS channel when the numbers of transmit and receive paths are both equal to 1. Consequently, the channel between the BS and the $k$-th target is given as 
	\begin{align}
		\mathbf{h}_{d,k} = \sqrt{\alpha_{d,k}}\mathbf{a}_M(\rho_{B,k},\mathbf{b}),
	\end{align}
	where $\rho_{B,k}=\phi_{B,k}$, $\phi_{B,k}$ is the AoD of the $k$-th target with respect to the BS, $\mathbf{a}_M(\rho_{B,k},\mathbf{b})=[1, e^{j\frac{2\pi}{\lambda}\rho_{B,k}b_1}, ...,e^{j\frac{2\pi}{\lambda}\rho_{B,k}b_M}]^T$ is the steering vector of the BS, and $\alpha_{d,k}$ is the distance dependent free space path loss. Then, the channel between the $k$-th target and the MIRS is obtained by
	\begin{equation}
		\mathbf{h}_{r,k}(\mathbf{t}) = \sqrt{\alpha_{r,k}}\mathbf{a}_N(\boldsymbol{\rho}_{I,k},\mathbf{t}),
	\end{equation}
	where $\boldsymbol{\rho}_{I,k} = [\sin \theta_{I,k}\cos \phi_{I,k},\cos \theta_{I,k}]^T$, $\theta_{I,k}$ and $\phi_{I,k}$ are the elevation and azimuth AoD of the $k$-th target with respect to the MIRS. The steering vector is then given as $\mathbf{a}_N(\boldsymbol{\rho}_{I,k},\mathbf{t})=\left[e^{\jmath\frac{2\pi}{\lambda}\boldsymbol{\rho}_{I,k}^T\mathbf{t}_1},\dots,e^{\jmath\frac{2\pi}{\lambda}\boldsymbol{\rho}_{I,k}^T\mathbf{t}_N}\right]^T$. Similar to \eqref{Ch1}, the channel between the BS and the target $k$ -th can be denoted as
	\begin{equation}
		\label{Sh1}
		\mathbf{h}^H_{S,k}(\mathbf{t},\boldsymbol{\phi}) =\mathbf{h}^H_{d,k} + \boldsymbol{\phi}^H\operatorname{diag}(\mathbf{h}^H_{r,k}(\mathbf{t}))\mathbf{G}, \forall k \in \mathcal{K}_t.
	\end{equation}

	\subsection{Channel Modeling for MIRS With Array-Wise Control}
	
	We then introduce the channel modeling when array-wise control is implemented for the MIRS. For convenience, we consider square arrays composed of a perfect square number of elements, with each row (or column) containing $a$ elements, resulting in a total of $a^2$ elements in each array. The spacing between adjacent elements is set to half the wavelength $\lambda$. Some examples are given in Fig. 2. We denote the position of the $n$-th movable array as $\mathbf{u}_n$, which is the coordinate of its geometric center, then the position of the element in the $i$-th row, $j$-th column within the array can be denoted as 
	\begin{align}
		\mathbf{u}_{n,i,j} = \mathbf{u}_n + \lambda(\frac{2j-a-1}{4},\frac{-2i+a+1}{4}).
	\end{align} 
	For a total element number of $N$ and $a^2$ elements in each movable array, the set of the array positions is given as $\mathbf{u}=[\mathbf{u}^T_1,\dots,\mathbf{u}^T_{N/a^2}]^T\in\mathbb{R}^{2N/a^2}$, and the positions of all the elements can be obtained as
	\begin{align}
		&\mathbf{u}'(\mathbf{u})=[(\mathbf{u}'_1)^T,\dots,(\mathbf{u}'_N)^T]\nonumber&\\&= [\mathbf{u}^T_{1,1,1},\!\dots\!,\mathbf{u}^T_{1,a,a},\!\dots\!,\mathbf{u}^T_{N/a^2,1,1},\!\dots\!,\mathbf{u}^T_{N/a^2,a,a}]\!\in\!\mathbb{R}^{2N}.
	\end{align} 
	
	With the movable arrays, the FRMs of the MIRS for the arrival and reflecting paths are obtained as $\mathbf{F}_a\bigl(\mathbf{u}'(\mathbf{u})\bigr)$ and $\mathbf{F}_d\bigl(\mathbf{u}'(\mathbf{u})\bigr)$, respectively. Then, the BS-IRS channel can be obtained as
	\begin{align}
		\mathbf{G}(\mathbf{u})=\mathbf{F}^H_a\bigl(\mathbf{u}'(\mathbf{u})\bigr)\boldsymbol{\Sigma}_{G}\mathbf{G}_t.
	\end{align}
	\begin{figure}[t]
		\centering{\includegraphics[width=0.7\columnwidth]{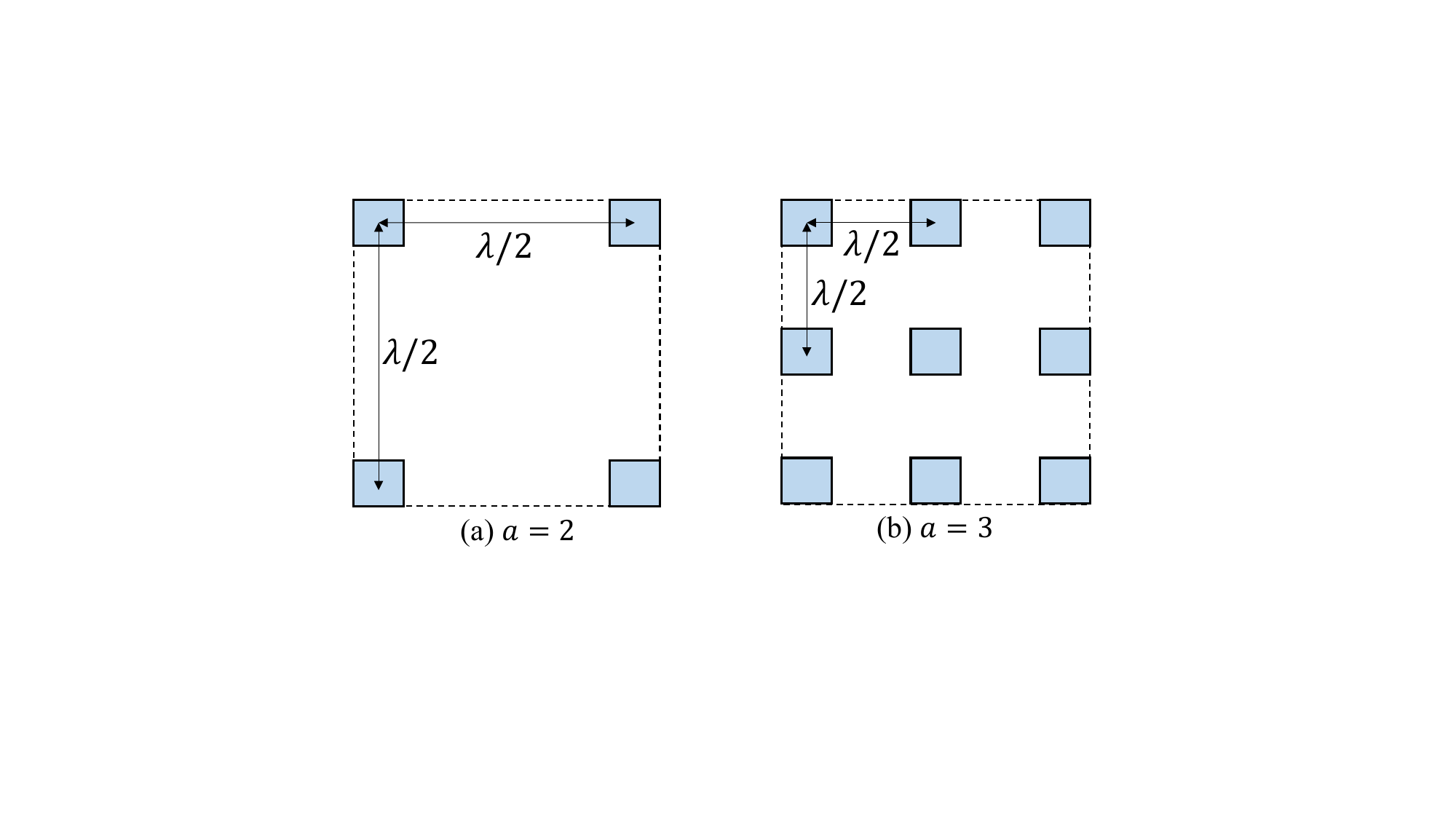}}
		\caption{Examples of the ME arrays with (a) $a=2$ and (b) $a=3$.}
	\end{figure}
	For the communication part, the channels between the MIRS and CUs are given as
	\begin{equation}
		\mathbf{f}_k(\mathbf{u}) = \mathbf{F}_d\bigl(\mathbf{u}'(\mathbf{u})\bigr)^H\mathbf{\Sigma}_{\mathrm{f},k}\mathbf{1}, \forall k \in\mathcal{K}_c.
	\end{equation}
	
	For the sensing part, the channels between the MIRS and targets are
	\begin{equation}
		\mathbf{h}_{r,k}(\mathbf{u}) = \sqrt{\alpha_{r,k}}\mathbf{a}_N\bigl(\boldsymbol{\rho}_{I,k},\mathbf{u}'(\mathbf{u})\bigr), \forall k \in \mathcal{K}_t.
	\end{equation}
	
	Based on the above, with the array-wise control mode, the channels between the BS and the $k$-th CU and $k$-th target are given as
	\begin{equation}
		\label{Ch2}
		\mathbf{h}^H_{C,k}(\mathbf{u},\boldsymbol{\phi}) =\boldsymbol{\phi}^H\operatorname{diag}(\mathbf{f}^H_{k}(\mathbf{u}))\mathbf{G}(\mathbf{u}), \forall k \!\in\! \mathcal{K}_c
	\end{equation}
	and 
	\begin{equation}
		\label{Sh2}
		\mathbf{h}^H_{S,k}(\mathbf{u},\boldsymbol{\phi}) =\mathbf{h}^H_{d,k} + \boldsymbol{\phi}^H\operatorname{diag}(\mathbf{h}^H_{r,k}(\mathbf{u}))\mathbf{G}, \forall k \in \mathcal{K}_t,
	\end{equation}
	respectively. Compared to the element-wise control scheme, the proposed array-wise schemes manipulate the element positions of the MIRS by controlling position variable with lower size. This approach reduces computational complexity at the cost of some available DoFs, thereby offering a practical trade-off between performance and computational overhead for practical applications.
	
	To investigate the theoretical performance of the MIRS-aided multi-user multi-target ISAC systems, we assume that the perfect CSI for all the channels is known in the following sections, given existing channel estimation (CE) methods for MA-aided wireless systems \cite{CE1,CE2}. While considering the potential random multipath interference in wireless environments, the CSI in practice may be imperfect \cite{CSI}. We will demonstrate the effectiveness of the proposed method under imperfect CSI by the simulation results in Section V.

	\subsection{Signal Model}
	The BS provides downlink communication for the $K_c$ CUs and performs radar sensing for the $K_t$ targets simultaneously. For the monostatic ISAC, we design the dual-functional signal transmitted by the BS at the $t$-th time slot as
	\begin{align}
		\mathbf{x}[t]=\mathbf{W}_c\mathbf{s}_c[t]+\mathbf{W}_s\mathbf{s}_s[t]=\mathbf{W}\mathbf{s}[t],
	\end{align}
	where $\mathbf{s}_c[t]=\bigl[{s}_{c,1}[t],\dots,{s}_{c,K_c}[t]\bigr]\in\mathbb{C}^{K_c}$ is the communication symbol vector and  ${s}_{c,k}[t]$ is the communication symbol for the $k$-th CU. $\mathbf{s}_s[t]\in\mathbb{C}^{K_t}$ denotes the radar sensing waveform transmitted at the $t$-th time slot. $\mathbf{W}_c\in\mathbb{C}^{M\times K_c}$ and $\mathbf{W}_s\in\mathbb{C}^{M\times K_t}$ are the beamforming matrices for the communication symbols and radar sensing beams, respectively. Then, the combined beamforming matrix and signal symbol are given as $\mathbf{W}=[\mathbf{W}_c\ \mathbf{W}_s]=[\mathbf{w}_1,\dots,\mathbf{w}_{K_c},\mathbf{w}_{K_c+1},\dots,\mathbf{w}_{K_c+K_t}]\in\mathbb{C}^{M\times (K_c+K_t)}$ and $\mathbf{s}[t]=[\mathbf{s}^T_c[t]\  \mathbf{s}^T_s[t]]^T\in\mathbb{C}^{K_c+K_t}$, respectively. We assume that the communication symbols are independent and
	identically distributed (i.i.d.) random variables with zero mean
	and unit variance, and the radar waveforms are generated by pseudo
	random coding, which leads to the condition $\mathbb{E}\{\mathbf{s}[t]\mathbf{s}^H[t]\}=\mathbf{I}_{K_c+K_t}$. In each ISAC period, a total number of $T$ combined symbols are assumed to be transmitted. The total symbol matrix is denoted as $\mathbf{S}=[\mathbf{s}[1],\dots,\mathbf{s}[T]]\in\mathbb{C}^{(K_c+K_t)\times T}$. With sufficient large $T$, we have $\mathbf{S}\mathbf{S}^H\approx T\mathbb{E}\{\mathbf{s}[t]\mathbf{s}^H[t]\}=T\mathbf{I}_{K_c+K_t}$. Then, the transmitted dual-functional signal is given as $\mathbf{X}=\mathbf{W}\mathbf{S}$.
	
	For the far-field radar sensing, the echo signal received by the BS from different targets may have different delays. Suppose that the echo signals received from the $k$-th target delays $r_k$ time slots compared with the transmitted signal, and the maximum number of the delayed time slots is $R$. In this work, we consider the target tracking task for radar sensing. In the initial target detection stage, the target positions of interest can be obtained by coarse sensing such as radar scanning. In the tracking stage, the target position obtained in previous sensing stage can be utilized. Thus, we assume that the positions and the corresponding time delays for different targets are known at the BS. To avoid the echo signal reflected by the farthest target in the current sensing period interfering with the echo signal reflected by the nearest target in the next period, the pulsed signal $[\mathbf{S}\ \mathbf{0}_{(K_c+K_t)\times R}]\in\mathbb{C}^{(K_c+K_t)\times (T+R)}$ is transmitted in each sensing period, where $T$ is typically designed much larger than $R$ to keep efficiency loss minimal \cite{SINR1,SINR2}. Then we utilize a shift matrix to describe the time delay of the signal for the $k$-th target as $\mathbf{J}_k\in\mathbb{R}^{T\times (T+R-1)}$, which is given by
	\begin{equation}
		\mathbf{J}_k[i,j]=
		\begin{cases}
			1& j-i=r_k-1\\
			0& \operatorname{otherwise}
		\end{cases}.
	\end{equation}
	
	Since we have the time delay information of the targets, we apply matched-filtering $(\mathbf{S}\mathbf{J}_k)^H$ on the received signal. Then, the received signal during one sensing period is obtained as
	\begin{align}
		\mathbf{Y}_S&=\sum_{k\in\mathcal{K}_t}\left[\alpha_k\mathbf{H}_{S,k}\mathbf{W}\mathbf{S}\mathbf{J}_k(\mathbf{S}\mathbf{J}_k)^H+\mathbf{N}_{S,k}(\mathbf{S}\mathbf{J}_k)^H\right]\nonumber\\&=\sum_{k\in\mathcal{K}_t}\left[\alpha_k\mathbf{H}_{S,k}\mathbf{W}\mathbf{S}\mathbf{S}^H+\mathbf{N}_{S,k}\mathbf{J}_k^H\mathbf{S}^H\right],
	\end{align}
	where $\alpha_k$ is the radar cross section (RCS) of the $k$-th target, $\mathbf{H}_{S,k}=(\mathbf{h}^H_{S,k})^T\mathbf{h}^H_{S,k}=\mathbf{h}^*_{S,k}\mathbf{h}^H_{S,k}$ is the cascaded echo channel for the $k$-th target. $\mathbf{N}_{S,k}=\bigl[\mathbf{n}_{S,k}[1],\dots,\mathbf{n}_{S,k}[T+R-1]\bigr]$ is the noise for the $k$-th target at the BS, which contains the uncorrelated interference from the environment.  $\mathbf{n}_{S,k}[t]\sim{\mathcal{CN}(0,\mathbf{I}_{M}\sigma_S^2)}$, and $\sigma_{S}^2$ is the noise power. 
	
	For the communications part, the signal received by the $k$-th CU can be obtained as
	\begin{align}
		\label{comsig}
		\mathbf{y}_k=\mathbf{h}^H_{C,k}\mathbf{X}+\mathbf{n}_k=\mathbf{h}^H_{C,k}\mathbf{W}\mathbf{S}+\mathbf{n}_k,
	\end{align}
	where $\mathbf{n}_k\sim{\mathcal{CN}(0,\mathbf{I}_L\sigma_{c,k}^2)}$ is the additive white Gaussian noise (AWGN) at the $k$-th CU with noise power $\sigma_{c,k}^2$. 
	
	\section{Performance Metrics and Problem Formulation}
	In this section, we introduce the metrics for evaluating the performance of the MIRS-aided multi-user multi-target ISAC systems and formulate the problem to be tackled.
	\subsection{Achievable Rate for Multi-User Communications}
	The achievable rate is adopted as the metric to evaluate whether the QoS of each CU is satisfied. Based on the signal model \eqref{comsig}, the achievable rate of the $k$-th CU is $R_k = \log\bigl(1+\gamma_{C,k}\bigr)$, where
	\begin{align}
		\gamma_{C,k}=\frac{\lvert\mathbf{h}_{C,k}^H\mathbf{w}_k\rvert^2}{\sum_{j\neq k}^{K_c+K_t}\lvert\mathbf{h}_{C,k}^H\mathbf{w}_j\rvert^2 + \sigma^2_{c,k}}
	\end{align} 
	is the SINR for the $k$-th CU.
	
	\subsection{SINR for Target Detection in Radar Sensing}
	Considering target tracking as the sensing task, the probability of detection for a target is positively proportional to the radar SINR \cite{IRS3}, which is considered as the sensing metric in this work. The radar SINR is obtained as follows. First, the received echo signal is vectorized. Denote $\mathbf{y}_S=\operatorname{vec}(\mathbf{Y}_s)$, $\mathbf{w}=\operatorname{vec}(\mathbf{W})$, and $\mathbf{n}_S=\sum_{k\in\mathcal{K}_t}\operatorname{vec}(\mathbf{N}_{S,k}\mathbf{J}_k^H\mathbf{S}^H)$, the vectorized echo signal is obtained as 
	\begin{align}
		\mathbf{y}_S=\sum_{k\in \mathcal{K}_t}\left[\alpha_k\left(\mathbf{S}\mathbf{S}^H\otimes\mathbf{H}_{S,k}\right)\mathbf{w}+\mathbf{n}_S\right].
	\end{align}
	Then, we apply receive filters to process the echo signal. Denote the receive filter for the $k$-th target and the filter matrix containing all the filters as $\mathbf{v}_k\in\mathbb{C}^{M(K_c+K_t)}$ and $\mathbf{V}=[\mathbf{v}_1,\dots,\mathbf{v}_{K_t}]\in\mathbb{C}^{M(K_c+K_t)\times K_t}$, respectively. Then, for the $k$-th target, the filtered signal is obtained as
	\begin{align}
		\mathbf{v}_k^H\mathbf{y}_S&=\alpha_k\mathbf{v}_k^H\left(\mathbf{S}\mathbf{S}^H\otimes\mathbf{H}_{S,k}\right)\mathbf{w}\nonumber\\&+\mathbf{v}_k^H\!\!\!\sum_{j\in \mathcal{K}_t,j\neq k}\!\!\!\alpha_j\left(\mathbf{S}\mathbf{S}^H\otimes\mathbf{H}_{S,j}\right)\mathbf{w}+\mathbf{v}_k^H\mathbf{n}_S.
	\end{align}
	Thus, the output SINR for the $k$-th target is given as 
	\begin{align}
		\gamma_{S,k}\!=\!\frac{\alpha_k^2\mathbb{E}\{\lvert\mathbf{v}_k^H\left(\mathbf{S}\mathbf{S}^H\otimes\mathbf{H}_{S,k}\right)\mathbf{w}\rvert^2\}}{\sum_{j\neq k}\!\alpha_j^2\mathbb{E}\{\lvert\mathbf{v}_k^H\left(\mathbf{S}\mathbf{S}^H\otimes\mathbf{H}_{S,j}\right)\mathbf{w}\rvert^2\}\!+\!K_tT\sigma_S^2\mathbf{v}_k^H\mathbf{v}_k}.
	\end{align}
	
	In this work, to characterize the fundamental system performance, we assume that the number of symbols transmitted during one sensing period is large enough to satisfy $\mathbf{S}\mathbf{S}^H = T\mathbf{I}_{K_c+K_t}$. Then, the output SINR for the $k$-th target can be obtained as
	\begin{align}
		\gamma_{S,k}\!=\!\frac{T\alpha_k^2\lvert\mathbf{v}_k^H\left(\mathbf{I}_{K_c+K_t}\otimes\mathbf{H}_{S,k}\right)\mathbf{w}\rvert^2}{T\sum_{j\neq k}\alpha_j^2\lvert\mathbf{v}_k^H\left(\mathbf{I}_{K_c+K_t}\otimes\mathbf{H}_{S,j}\right)\mathbf{w}\rvert^2+K_t\sigma_S^2\mathbf{v}_k^H\mathbf{v}_k}.
	\end{align}

	\subsection{Problem Formulation}
	In the ISAC systems, the minimum communication rate for CUs must be guaranteed to satisfy QoS requirements. At the same time, target tracking necessitates a minimum SINR to ensure the desired probability of detection. In this work, we investigate whether applying MIRS can enhance the channel environment, thereby enabling the required minimum communication rate and sensing SINR to be achieved under reduced transmit power. Since element-wise control can be regarded as a special case of array-wise control when $a=1$, we formulate the problem based on the array-wise control scheme. Denote the set of controllable positions as $\mathcal{N}\triangleq\{1,\dots,\frac{N}{a^2}\}$, the problem is formulated as 
	\begin{subequations}
		\label{Q}
		\begin{IEEEeqnarray}{r,l}			
			$$\underset{\mathbf{W},\mathbf{V},\boldsymbol{\phi},\mathbf{u}}{\min}$$&{\ \lVert\mathbf{W}\rVert_F^2}\\
			$$\operatorname{s.t.}$$ 
			&\ R_k \geq \Gamma_k,  \forall k \in \mathcal{K}_c\label{Cr},\\
			&\ \gamma_{S,k}\geq \chi_k, \forall k \in \mathcal{K}_t, \label{Cs}\\
			&\ \lVert\mathbf{v}_k\rVert_2^2=1, \forall k \in\mathcal{K}_t, \label{Cv}\\\
			&\ \lvert\phi_n\rvert=1, \forall n \in \mathcal{N}, \label{Cphi}\\
			&\ \mathbf{u}_n \in \mathcal{C}, \forall n \in \mathcal{N}, \label{Cregion}\\
			&\ \lVert \mathbf{u}_n-\mathbf{u}_{n'}\rVert \geq d_{\min}, \forall n\neq n',n,n'\in\mathcal{N}. \label{Ct}
		\end{IEEEeqnarray}
	\end{subequations}
	
	The objective function represents the total transmit power required at the BS for the beamforming. Equations \eqref{Cr} and \eqref{Cs} denote the minimum communication rate and sensing SINR constraint, respectively, where $\Gamma_k$ and $\chi_k$ are the required minimum rate for the $k$-th CU and the minimum SINR for the $k$-th target. Equation \eqref{Cv} constrains the power of the receive filters, Equation \eqref{Cphi} is the constant modulus constraints (CMC) for the IRS phase shifts, Equations \eqref{Cregion} and \eqref{Ct} represent that the elements of the MIRS should be confined in the given region $\mathcal{C}$ and keep a minimum distance of $d_{\min}$ to avoid the antenna coupling effect, where $d_{\min}=\frac{\sqrt{2}(a-1)+2}{4}\lambda$ is geometrically chosen to ensure that the distance between any two elements from the two arrays is larger than half a wavelength. 
	
	It can be observed that problem \eqref{Q} is highly non-convex due to the non-convex constraints. Although existing techniques could be utilized to obtain a feasible solution for the variables separately, such as the semidefinite relaxation (SDR)-based algorithm for the beamforming design \cite{MAIRSISAC1}, the joint optimization of the variables cannot be handled well. To tackle the problem, we propose the PRMO method to find feasible solutions of the problem by jointly optimizing the variables in the following sections.
	
	\section{PRMO For The Joint Beamforming and Position Optimization}
	The PRMO method is developed in this section to solve \eqref{Q}. The main steps are as follows: 1) The problem is reformulated over a constructed PRMS based on several Riemannian manifold spaces; 2) The problem is transformed as an unconstrained Riemannian problem via penalty based method and smoothing technique; 3) The exact penalty method is exploited to update the penalty weight and the RBFGS algorithm is utilized for updating the variables and finding a feasible solution.
	
	\subsection{Construction of the Solution Space}
	A Riemannian manifold is a smooth manifold, in where every point can be linearized locally as a tangent space. Each tangent space is equipped with its own smoothly varying inner product, thereby providing a local Euclidean structure at every point. This Riemannian structure enables the definition of fundamental differential geometric concepts such as gradients and Hessians on the manifold, facilitating the extension of optimization algorithms \cite{rmo}. We observe that the beamforming matrix $\mathbf{W}$ is constrained in Euclidean complex space, which is also a basic Riemannian manifold space. Besides, the power constraint of the receive filters \eqref{Cv} and the CMC of IRS reflection coefficient \eqref{Cphi} could be interpreted as confining the variables within the Riemannian complex oblique manifold (COM) and complex circle manifold (CCM), which are standard Riemannian manifold spaces and can be defined as that given in \cite{rmo} as
	\begin{equation}
		\mathcal{M}_{\mathbf{V}}=\{\mathbf{V}\in \mathbb{C}^{M(K_c+K_t)\times K_t} \mid \operatorname{diag}(\mathbf{V}^H\mathbf{V})=\mathbf{1}\}
	\end{equation}
	and
	\begin{flalign}	
		\mathcal{M}_{\boldsymbol{\phi}}=\{\boldsymbol{\phi}\in \mathbb{C}^{N} \mid \lvert \phi_n\rvert=1, \forall n \in \mathcal{N}\},
	\end{flalign} 	
	respectively. As given in \cite{rmo}, the COM and CCM can be linearized locally around every point as tangent spaces, which are given as
	\begin{flalign}	
		{\rm T}_\mathbf{V}\mathcal{M}_\mathbf{V} \!=\!\bigl\{\boldsymbol{\zeta}_\mathbf{V}\!\mid\! \boldsymbol{\zeta}_\mathbf{V}\!\in\! \mathbb{C}^{M(K_c+K_t)\times K_t}, \operatorname{diag}(\mathbf{V}^H\boldsymbol{\zeta}_\mathbf{V})=\mathbf{0}\bigr\}
	\end{flalign}
	and
	\begin{flalign}	
		{\rm T}_{\boldsymbol{\phi}}\mathcal{M}_{\boldsymbol{\phi}} \!=\!\bigl\{\boldsymbol{\zeta}_{\boldsymbol{\phi}}\!\mid\! \boldsymbol{\zeta}_{\boldsymbol{\phi}} \in \mathbb{C}^{N}, \Re(\boldsymbol{\zeta}^*_{\boldsymbol{\phi}}\odot\boldsymbol{\phi})=\mathbf{0}\bigr\},
	\end{flalign}
	respectively. Then, to tackle the constraints \eqref{Cregion} that confine the positions within the confined region, we introduce an auxiliary variable $\tilde{\mathbf{u}}=[\tilde{\mathbf{u}}^T_1,\dots,\tilde{\mathbf{u}}^T_{{N}/{a^2}}]^T=[\tilde{u}_1,\dots,\tilde{u}_{2N/a^2}]^T\in\mathbb{R}^{2N/a^2}$. Then, exploiting the sigmoid function, which is given as $\operatorname{sig}(x)=1/(1+e^{-x})\in(0,1)$, we define the projection functions on $\tilde{\mathbf{u}}$ as
	\begin{align}
		\label{projt}
		\mathbf{u}=\boldsymbol{p}(\tilde{\mathbf{u}}) = A\mathbf{sig}(\tilde{\mathbf{u}}),
	\end{align}
	where $\mathbf{sig}(\tilde{\mathbf{u}})=[\operatorname{sig}(\tilde{u}_1),\dots,\operatorname{sig}(\tilde{u}_{2N/a^2})]^T$. Then, problem \eqref{Q} is equivalent to 
	\begin{subequations}
		\label{Q2}
		\begin{IEEEeqnarray}{r,l}			
			$$\underset{\mathbf{W},\mathbf{V},\boldsymbol{\phi},\tilde{\mathbf{u}}}{\min}$$&{\ \lVert\mathbf{W}\rVert_F^2}\\
			$$\operatorname{s.t.}$$ 
			&\ \mathbf{W}\!\in\!\mathbb{C}^{M\!\times\!(K_c+K_t)}, \mathbf{V}\!\in\!\mathcal{M}_{\mathbf{V}}, \boldsymbol{\phi}\!\in\! \mathcal{M}_{\boldsymbol{\phi}}, \tilde{\mathbf{u}}\!\in\!\mathbb{R}^{\frac{2N}{a^2}}\!,\ \ \ \ \ \ \\
			&\ R_k \geq \Gamma_k,  \forall k \in \mathcal{K}_c \label{ineqR}\\
			&\ \gamma_{S,k}\geq \chi_k, \forall k \in \mathcal{K}_t, \label{ineqG}\\
			&\ \lVert \boldsymbol{p}(\tilde{\mathbf{u}}_n)\!-\!\boldsymbol{p}(\tilde{\mathbf{u}}_{n'})\rVert\!\geq\!d_{\min},\! \forall n\!\neq\! n',n,n'\!\!\in\!\mathcal{N}. \label{ineqP}
		\end{IEEEeqnarray}
	\end{subequations}
	
	We can observe that the variables are all confined in smooth Riemannian submanifolds. Instead of handling these constraints separately, we reformulate the problem over a PRMS combining the submanifolds, which allows all variables to be jointly and simultaneously optimized in a unified geometric space while naturally preserving their intrinsic constraints. Define a produce variable as $\mathbf{X} = (\mathbf{W},\mathbf{V}, \boldsymbol{\phi}, \tilde{\mathbf{u}})$. By combining the submanifolds, a PRMS $\mathcal{M}$ serving as the solution space for $\mathbf{X}$ can be constructed as
	\begin{align}	
		\mathcal{M} =\{\mathbf{X}=(\mathbf{W},\mathbf{V}, \boldsymbol{\phi}, \tilde{\mathbf{u}}) \mid \mathbf{W}&\in\mathbb{C}^{M\times(K_c+K_t)},\mathbf{V}\in\mathcal{M}_{\mathbf{V}}, \nonumber\\&\boldsymbol{\phi}\in\mathcal{M}_{\boldsymbol{\phi}},\tilde{\mathbf{u}}\in\mathbb{R}^{\frac{2N}{a^2}}\}.
	\end{align}
	The tangent space of the point $\mathbf{X}$ over $\mathcal{M}$ is then obtained as
	\begin{align}	
		{\rm T}_{\mathbf{X}} \mathcal{M} =&\bigl\{\boldsymbol{\zeta}_{\mathbf{X}} = (\boldsymbol{\zeta}_\mathbf{W},\boldsymbol{\zeta}_\mathbf{V}, \boldsymbol{\zeta}_{\boldsymbol{\phi}},\boldsymbol{\zeta}_{\tilde{\mathbf{u}}})\mid\boldsymbol{\zeta}_\mathbf{W}\in \mathbb{C}^{M\times(K_c+K_t)}, \nonumber\\& \boldsymbol{\zeta}_\mathbf{V}\in {\rm T}_\mathbf{V}\mathcal{M}_\mathbf{V},\boldsymbol{\zeta}_{\boldsymbol{\phi}}\in{\rm T}_{\boldsymbol{\phi}}\mathcal{M}_{\boldsymbol{\phi}},\boldsymbol{\zeta}_{\tilde{\mathbf{u}}}\in\mathbb{R}^{\frac{2N}{a^2}}\bigr\}.
	\end{align}
	
	For various $\mathbf{X}\in\mathcal{M}$, the tangent space is different. To obtain different inner products for each point, we equip $\mathcal{M}$ with the product of Euclidean inner products as the Riemannian metric, which varies smoothly with the points and is defined as
	\begin{align}	&\langle\boldsymbol{\zeta}_{\mathbf{X}},\boldsymbol{\zeta}'_{\mathbf{X}}\rangle=\langle(\boldsymbol{\zeta}_\mathbf{W},\boldsymbol{\zeta}_\mathbf{V}, \boldsymbol{\zeta}_{\boldsymbol{\phi}},\boldsymbol{\zeta}_{\tilde{\mathbf{u}}}),(\boldsymbol{\zeta}'_\mathbf{W},\boldsymbol{\zeta}'_\mathbf{V}, \boldsymbol{\zeta}'_{\boldsymbol{\phi}},\boldsymbol{\zeta}'_{\tilde{\mathbf{u}}})\rangle\nonumber\\ =&\Re\left(\operatorname{Tr}(\boldsymbol{\zeta}^H_{\mathbf{W}}\boldsymbol{\zeta}'_{\mathbf{W}})\right)  +\Re\left(\operatorname{Tr}(\boldsymbol{\zeta}^H_{\mathbf{V}}\boldsymbol{\zeta}'_{\mathbf{V}})\right) + \Re(\boldsymbol{\zeta}_{\boldsymbol{\phi}}^H\boldsymbol{\zeta}'_{\boldsymbol{\phi}}) + \boldsymbol{\zeta}_{\tilde{\mathbf{u}}}^T\boldsymbol{\zeta}_{\tilde{\mathbf{u}}},
	\end{align}
	and the norm for tangent vectors is obtained as $\lVert \boldsymbol{\zeta}_{\mathbf{X}}\rVert = \sqrt{\langle\boldsymbol{\zeta}_{\mathbf{X}},\boldsymbol{\zeta}_{\mathbf{X}}\rangle}$.

	\subsection{Problem Reformulation via Penalty Method}
	The minimum communication rate constraint \eqref{ineqR} and sensing SINR constraint \eqref{ineqG} could be written as
	\begin{equation}
		h_k(\mathbf{X}) = \Gamma_k - r_k \leq 0,\forall k \in \mathcal{K}_c
	\end{equation}
	and
	\begin{equation}
		h_k(\mathbf{X}) = \chi_k - \gamma_{S,k} \leq 0,\forall k \in \mathcal{K}_t,
	\end{equation}
	respectively. The minimum distance constraint \eqref{ineqP} can be rewritten as
	\begin{align}
		h_{n,n'}(\mathbf{X})=d_{\min} -\lVert \boldsymbol{p}(\tilde{\mathbf{u}}_n)-&\boldsymbol{p}(\tilde{\mathbf{u}}_{n'})\rVert \leq 0, \nonumber\\&\forall n,n' \in \mathcal{N}, n \neq n'.
	\end{align} 
	Define the set of the minimum distance constraints as $\mathcal{D} \triangleq \{(i,i')\mid \forall i,i'\in\mathcal{N}, i\neq i'\}$, the set of all the inequality constraints could be denoted as $\mathcal{I} = \mathcal{D} \cup \mathcal{K}_t \cup \mathcal{K}_c$. Then, problem \eqref{Q2} can be rewritten as
	\begin{subequations}
		\label{Q3}
		\begin{IEEEeqnarray}{r,l}			
			$$\underset{\mathbf{X}}{\min}$$&{\ f(\mathbf{X})=\lVert\mathbf{W}\rVert_F^2}\\
			$$\operatorname{s.t.}$$ &\ \mathbf{X} \in \mathcal{M},\\
			&\ h_i(\mathbf{X}) \leq 0, \forall i\in\mathcal{I}. \label{ineq}
		\end{IEEEeqnarray}
	\end{subequations}
	
	We then utilize the penalty method to tackle the inequality constraints, which augments the objective function with a weighted penalty term $\rho \sum_{i \in \mathcal{I}} \max\{0, h_i(\mathbf{X})\}$ to penalize violations of the constraints, where $\rho \geq 0$ denotes the penalty weight \cite{penalty, ep}. It is well established that exact satisfaction of the constraints can be attained with a finite penalty weight in the Euclidean case \cite{penalty}, and the result has been extended to the Riemannian case as well \cite{rep}. However, the penalty is intractable since it is nonsmooth and has non-differential discontinuities. To tackle it, we utilize the linear-quadratic function \cite{lq}, which is smooth and can be given by $\max\{0,x\}\approx\mathcal{P}(x,u)$, where
	\begin{equation}
		\mathcal{P}(x,u)=
		\begin{cases}
			0& x\leq 0\\
			\frac{x^2}{2u}& 0<x\leq u\\
			x-\frac{u}{2}& x > u
		\end{cases}
	\end{equation}
	and $u\geq0$ is the smoothing parameter. Typically, a higher approximation accuracy is achieved with lower $u$. Based on the above, \eqref{Q2} can be converted as
	\begin{subequations}
		\label{Qs}
		\begin{IEEEeqnarray}{r,l}			
			$$\underset{\mathbf{X}}{\min}$$&{\  g(\mathbf{X})=f(\mathbf{X})+\rho\sum_{i\in \mathcal{I}}\mathcal{P}(h_i(\mathbf{X},u))} \\
			$$\operatorname{s.t.}$$ &\ \mathbf{X}\in\mathcal{M}.
		\end{IEEEeqnarray}
	\end{subequations}
	
	With proper values of $\rho$ and $u$, a feasible solution of \eqref{Q2} can be found by finding an optimum of \eqref{Qs}. However, \eqref{Qs} is non-convex over $\mathcal{M}$. In the next subsections, we apply the RBFGS algorithm over $\mathcal{M}$ to solve \eqref{Qs}. To improve computation efficiency, we then utilize the exact penalty method to find proper values of $\rho$ and $u$ that ensure the inequality constraints.
	
	\subsection{RBFGS Algorithm for the Unconstrained Problem}
	
	The BFGS algorithm is a quasi-Newton optimization method that iteratively updates an approximation of the inverse Hessian matrix to guide the search for a local optimum \cite{rho}. The RBFGS algorithm extends the classical BFGS method to optimization problems defined on Riemannian manifolds while enabling efficient optimization on curved geometries\cite{rbfgs}. In this section, we extend the RBFGS algorithm over the constructed PRMS $\mathcal{M}$ to tackle problem \eqref{Qs}.
	
	\subsubsection{Riemannian gradient over $\mathcal{M}$}
	We first derive the Riemannian gradient of the objective function with respect to the product variable $\mathbf{X}$ in this section. For the Euclidean gradients with respect to the beamforming matrix $\mathbf{W}$, receive filters $\mathbf{V}$, IRS reflection coefficient $\boldsymbol{\phi}$, and positions $\tilde{\mathbf{u}}$, we have
	\begin{equation}
		\nabla_{\mathbf{W}^*} g(\mathbf{X})=2\mathbf{W}+\rho\sum_{i\in\mathcal{K}_t\cup\mathcal{K}_c}\nabla_{\mathbf{W}^*} \mathcal{P}(h_i(\mathbf{X}),u),
	\end{equation}
	\begin{equation}
		\nabla_{\mathbf{V}^*} g(\mathbf{X})=\rho\sum_{i\in\mathcal{K}_t\cup\mathcal{K}_c}\nabla_{\mathbf{V}^*} \mathcal{P}(h_i(\mathbf{X}),u),
	\end{equation}
	\begin{equation}
		\nabla_{\boldsymbol{\phi}^*} g(\mathbf{X})=\rho\sum_{i\in\mathcal{K}_t\cup\mathcal{K}_c}\nabla_{\boldsymbol{\phi}^*} \mathcal{P}(h_i(\mathbf{X}),u),
	\end{equation}
	and
	\begin{equation}
		\nabla_{\tilde{\mathbf{u}}} g(\mathbf{X})=\rho\sum_{i\in\mathcal{D}}\nabla_{\tilde{\mathbf{u}}} \mathcal{P}(h_i(\mathbf{X}),u).
	\end{equation}
	For any $\boldsymbol{\upsilon}\in\{\mathbf{W},\mathbf{V}, \boldsymbol{\phi},\tilde{\mathbf{u}}\}$, we have
	\begin{equation}
		\nabla_{\boldsymbol{\upsilon}^*} \mathcal{P}(h_i(\mathbf{X}),u)\!\!=\!\!
		\begin{cases}
			\mathbf{0}& \!\!h_i(\mathbf{X})\leq 0\\
			\frac{h_i(\mathbf{X})}{u} \nabla_{\boldsymbol{\upsilon}^*}h_i(\mathbf{X})&\!\! 0<h_i(\mathbf{X})\leq u\\
			\nabla_{\boldsymbol{\upsilon}^*}h_i(\mathbf{X})& \!\!h_i(\mathbf{X}) > u
		\end{cases}\!\!,
	\end{equation}
	and for $h_i(\mathbf{X}),\forall i\in\mathcal{K}_t\cup\mathcal{K}_c$, we have
	\begin{equation}
		\nabla_{\boldsymbol{\upsilon}^*}h_i(\mathbf{X})= -\frac{\partial \gamma_{S,i}}{\partial \boldsymbol{\upsilon}^*},\forall i \in \mathcal{K}_t,
	\end{equation}
	\begin{equation}
		\nabla_{\boldsymbol{\upsilon}^*}h_i(\mathbf{X})= -\frac{\partial r_i}{\partial {\boldsymbol{\upsilon}^*}}, \forall i \in \mathcal{K}_c,
	\end{equation}
	then for $h_i(\mathbf{X}),\forall i\in\mathcal{D}$, we first have
	\begin{align}
		&\frac{\partial h_i(\mathbf{X})}{\partial \mathbf{u}'_{n}} = \bigl((\mathbf{u}'_{n}-\mathbf{u}'_{n'})^T(\mathbf{u}'_{n}-\mathbf{u}'_{n'})\bigr)^{-1/2}(\mathbf{u}'_{n}-\mathbf{u}'_{n'}),\nonumber\\&
		\frac{\partial h_i(\mathbf{X})}{\partial \mathbf{u}'_{n'}} =\!-\! \bigl((\mathbf{u}'_{n}\!-\!\mathbf{u}'_{n'})^T(\mathbf{u}'_{n}\!-\!\mathbf{u}'_{n'})\bigr)^{-1/2}(\mathbf{u}'_{n}\!-\!\mathbf{u}'_{n'}),
	\end{align}
	which are the $n$-th and $n'$-th elements of $\partial h_i(\mathbf{X})/\partial \mathbf{u}', \forall i\in\mathcal{D}$, while other elements of which are $0$.
	Then the Euclidean gradient with respect to $\tilde{\mathbf{u}}$ based on the chain rule is
	\begin{equation}
		\nabla_{\tilde{\mathbf{u}}}h_i(\mathbf{X}) =\mathcal{D}^T_\mathbf{u}\mathbf{u}'(\mathbf{u})\frac{\partial h_i(\mathbf{X})}{\partial \mathbf{u}'}\!\odot\! A\mathbf{sig}(\tilde{\mathbf{u}})\!\odot\! (1-\mathbf{sig}(\tilde{\mathbf{u}})).
	\end{equation}
	
	The detailed closed-form expressions of the above partial derivatives are provided in Appendix A. Based on the Euclidean gradients, the Riemannian gradient of $g(\mathbf{X})$ at a point over $\mathcal{M}$ is obtained based on the orthogonal projection of the Euclidean gradient to the tangent space. Specifically, for the COM and CCM, the Riemannian gradients are obtained as
	\begin{align}	
		\operatorname{grad}_{\mathbf{V}}g(\mathbf{X}) =\nabla_{\mathbf{V}^*}g(\mathbf{X})- \mathbf{V}\Re\{\operatorname{ddiag}\bigl(\mathbf{V}^H\nabla_{\mathbf{V}^*}g(\mathbf{X})\bigr)\}
	\end{align}
	and
	\begin{align}	
		\operatorname{grad}_{\boldsymbol{\phi}}g(\mathbf{X}) =\nabla_{\boldsymbol{\phi}^*}g(\mathbf{X})- \Re\{\nabla_{\boldsymbol{\phi}^*} g(\mathbf{X})\odot\boldsymbol{\phi}^*\}\odot \boldsymbol{\phi},
	\end{align}
	respectively. Besides, we have $\operatorname{grad}_\mathbf{W}g(\mathbf{X})=\nabla_{\mathbf{W}^*}g(\mathbf{X})$ and $\operatorname{grad}_{\tilde{\mathbf{u}}}g(\mathbf{X})=\nabla_{\tilde{\mathbf{u}}}g(\mathbf{X})$ since $\mathbf{W}$ and $\tilde{\mathbf{u}}$ are confined in Euclidean spaces. Based on the above, the product Riemannian gradient with respect to $\mathbf{X}$ can be obtained as
	\begin{align}
		\label{rg}
		\operatorname{grad}_{\mathbf{X}}g(\mathbf{X}) =\big[\operatorname{grad}_{\mathbf{W}}&g(\mathbf{X}),\operatorname{grad}_{\mathbf{V}}g(\mathbf{X}),\nonumber\\&\operatorname{grad}_{\boldsymbol{\phi}}g(\mathbf{X}), \operatorname{grad}_{\tilde{\mathbf{u}}}g(\mathbf{X})\big].
	\end{align}
	
	\subsubsection{RBFGS algorithm over the PRMS}
	For a function over a Riemannian manifold space, an optimal second-order descent direction can be calculated via the Newton equation, which leverages curvature information from the Hessian and yields enhanced convergence and robustness performance \cite{rmo}. However, the Hessian is expensive to calculate and may not be positive definite. Instead, we resort to the quasi-Newton method of the RBFGS algorithm to obtain an approximation of the Hessian inverse for calculating the update direction \cite{penalty}. The update direction at a point over $\mathcal{M}$ is obtained as the product of the directions obtained via the RBFGS algorithm over each submanifold, which is given as 
	\begin{align}	
		\label{d}
		\mathbf{d}_{\mathbf{X}}&= (\mathbf{d}_\mathbf{W},\mathbf{d}_\mathbf{V},\mathbf{d}_{\boldsymbol{\phi}},\mathbf{d}_{\tilde{\mathbf{u}}}) = -\mathbf{H}_{\mathbf{X}}\operatorname{grad}_\mathbf{X}g(\mathbf{X})\nonumber\\& =- \bigl(\mathbf{H}_\mathbf{W}\operatorname{grad}_{\mathbf{W}}  g(\mathbf{X}),\mathbf{H}_\mathbf{V}\operatorname{grad}_{\mathbf{V}}  g(\mathbf{X}),\nonumber\\&\qquad\qquad \mathbf{H}_{\boldsymbol{\phi}}\operatorname{grad}_{\boldsymbol{\phi}}g(\mathbf{X}),\mathbf{H}_{\tilde{\mathbf{u}}}\operatorname{grad}_{\tilde{\mathbf{u}}}  g(\mathbf{X})\bigr),
	\end{align}
	where $\mathbf{H}_{\mathbf{X}}$ is the approximation to the inverse Hessian at $\mathbf{X}$. Updating the Hessian inverse approximations over $\mathcal{M}$ requires manifold-specific operations, including transport and retraction, to ensure that search directions and Hessian approximations remain within the tangent spaces. For a point $\mathbf{X}$ over $\mathcal{M}$ with an update direction $\mathbf{d}_\mathbf{X}$ and step size $\alpha$, the retraction operation for mapping the updated point from a tangent space back to the PRMS is defined as
	\begin{align}
		\label{ret}
		&\mathcal{R}_{\mathbf{X}}(\alpha \mathbf{d}_{\mathbf{X}}) =\nonumber\\&\bigl(\mathbf{W}+\alpha\mathbf{d}_{\mathbf{W}},(\mathbf{V}+\alpha\mathbf{d}_{\mathbf{V}})\operatorname{diag}(\lVert\mathbf{V}_{:1}\rVert^{-1}_2,\dots,\lVert\mathbf{V}_{:{K_t}}\rVert^{-1}_2),\nonumber\\&\quad(\boldsymbol{\phi}+\alpha\mathbf{d}_{\boldsymbol{\phi}}) \oslash \lvert\boldsymbol{\phi}+\alpha\mathbf{d}_{\boldsymbol{\phi}}\rvert,  \tilde{\mathbf{u}}+\alpha\mathbf{d}_{\tilde{\mathbf{u}}} \bigr).
	\end{align}
	To ensure that different gradients and search directions are compared in the same tangent space, the transport operation that coherently maps tangent vectors from other tangent spaces to the tangent space of a point $\mathbf{X}$ while preserving their geometric meaning is defined as 
	\begin{align}
		\label{tra}
		\mathcal{T}_{\mathbf{X}}(\mathbf{d}_{\mathbf{X}'})= \bigl(\mathbf{d}_{\mathbf{W}'},&\mathbf{d}_{\mathbf{V}'}-\mathbf{X}\Re\{\operatorname{diag}(\mathbf{V}^H\mathbf{d}_{\mathbf{V}'})\},\nonumber\\&\ \  \mathbf{d}_{\boldsymbol{\phi}'} - \Re\{\mathbf{d}_{\boldsymbol{\phi}'}^*\odot\boldsymbol{\phi}\}\odot \boldsymbol{\phi}, \mathbf{d}_{\tilde{\mathbf{u}}'} \bigr).
	\end{align}
	
	For the $(l+1)$-th iteration, the inverse Hessian approximation $\mathbf{H}_\mathbf{X}$ is obtained as the product of the inverse Hessian approximations in each submanifold and based on the medium variables, including
	\begin{equation}
		\label{s}
		\mathbf{S}^l_\mathbf{X} = (\mathbf{S}^l_\mathbf{W},\mathbf{S}^l_\mathbf{V},\mathbf{S}^l_{\boldsymbol{\phi}},\mathbf{S}_{\tilde{\mathbf{u}}})= \mathcal{T}_{\mathbf{X}^{l+1}}(\alpha^l\mathbf{d}_{\mathbf{X}}^l)
	\end{equation}
	and
	\begin{align}
		\label{y}
		\mathbf{Y}^l_\mathbf{X} \!&=\!(\mathbf{Y}^l_\mathbf{W},\mathbf{Y}^l_\mathbf{V},\mathbf{Y}^l_{\boldsymbol{\phi}},\mathbf{Y}_{\tilde{\mathbf{u}}})\!\nonumber\\&=\! \operatorname{grad}_{\mathbf{X}}g(\mathbf{X}^{l+1}) - \mathcal{T}_{\mathbf{X}^{l+1}}(\operatorname{grad}_{\mathbf{X}}g(\mathbf{X}^{l})).
	\end{align}
	To avoid the scaling of the RBFGS direction, the medium variables should be normalized, which can be implemented as $\mathbf{s}^l = \mathbf{s}_{\mathbf{X}}^l/\lVert\mathbf{s}_{\mathbf{X}}^l\rVert$ and $\mathbf{y}_{\mathbf{X}}^l = \mathbf{y}_{\mathbf{X}}^l/\lVert\mathbf{s}_{\mathbf{X}}^l\rVert$. Then, the inverse Hessian approximation with respect to $\mathbf{W}$, $\mathbf{V}$, $\boldsymbol{\phi}$, and $\tilde{\mathbf{u}}$ can be obtained. Take $\mathbf{W}$ as an example, it is calculated as
	\begin{align}
		\label{Hw}
		\mathbf{H}^{l+1}_\mathbf{W} =(\mathbf{Z}_{\mathbf{W}} ^l)^H\mathbf{H}_\mathbf{W}^{l}\mathbf{Z}_{\mathbf{W}}^l + \delta^l\mathbf{S}^l_{\mathbf{W}}(\mathbf{S}^l_{\mathbf{W}})^H,
	\end{align}
	where $\delta^l=1/\langle\mathbf{S}^l_{\mathbf{X}},\mathbf{Y}^l_{\mathbf{X}}\rangle$ and $\mathbf{Z}_{\mathbf{W}} ^l=\mathbf{I}-\delta^l\mathbf{S}^l_{\mathbf{W}}(\mathbf{Y}^l_{\mathbf{W}})^H$. $\mathbf{H}^{l+1}_\mathbf{V}$, $\mathbf{H}^{l+1}_{\boldsymbol{\phi}}$ and $\mathbf{H}^{l+1}_{\tilde{\mathbf{u}}}$ can be obtained similarly. Based on the above, the product inverse Hessian approximation $\mathbf{H}^{l+1}_{\mathbf{X}}$ can be obtained as
	\begin{equation}
		\mathbf{H}^{l+1}_{\mathbf{X}}=\left[\mathbf{H}^{l+1}_\mathbf{W},\mathbf{H}^{l+1}_\mathbf{V},\mathbf{H}^{l+1}_{\boldsymbol{\phi}},\mathbf{H}^{l+1}_{\tilde{\mathbf{u}}}\right].
	\end{equation}
	
	Instead of calculating \eqref{Hw} directly, an efficient way to calculate the inverse Hessian approximations is to utilize a limited memory to store limited medium variables obtained in the last iterations to compute the update direction, which reduces the order of computational complexity, avoids gradually dense approximations, and saves storage space \cite{penalty}. Define the limited memory as $\mathbf{M}$ with a size of $S$, no more than $S$ medium variable sets $\mathbf{M}_i = (\mathbf{S}^i_{\mathbf{X}},\mathbf{Y}^i_{\mathbf{X}},\delta^i), i\leq S$ obtained in the last iterations are stored for obtaining the update direction. Note that \eqref{Hw} have recursive properties, it can be expanded $m$ times as
	\begin{align}
		\label{HW}
		\mathbf{H}^{l}_\mathbf{W}&=\bigl(\mathbf{Z}^{l-m}_\mathbf{W}\mathbf{Z}^{l-m+1}_\mathbf{W}\dots\mathbf{Z}^{l-1}_\mathbf{W}\bigr)^H \mathbf{H}^{l-m}_\mathbf{W}\bigl(\mathbf{Z}^{l-m}_\mathbf{W}\dots\mathbf{Z}^{l-1}_\mathbf{W}\bigr) \nonumber\\&+ \delta^{l-m}\bigl(\mathbf{Z}^{l-m+1}_\mathbf{W}\mathbf{Z}^{l-m+2}_\mathbf{W}\dots\mathbf{Z}^{l-1}_\mathbf{W}\bigr)^H\mathbf{S}^{l-m}_\mathbf{W}\nonumber\\&\ \ \ (\mathbf{S}^{l-m}_\mathbf{W})^H\bigl(\mathbf{Z}^{l-m+1}_\mathbf{W}\dots\mathbf{Z}^{l-1}_\mathbf{W}\bigr)\nonumber\\&+\delta^{l-m+1}\bigl(\mathbf{Z}^{l-m+2}_\mathbf{W}\mathbf{Z}^{l-m+3}_\mathbf{W}\dots\mathbf{Z}^{l-1}_\mathbf{W}\bigr)^H\mathbf{S}^{l-m+1}_\mathbf{W}\nonumber\\&\ \ \  (\mathbf{S}^{l-m+1}_\mathbf{W})^H\bigl(\mathbf{Z}^{l-m+2}_\mathbf{W}\dots\mathbf{Z}^{l-1}_\mathbf{W}\bigr)\nonumber\\&+\dots+\delta^{l-1}\mathbf{S}^{l-1}_\mathbf{W}(\mathbf{S}^{l-1}_\mathbf{W})^H,
	\end{align}
	and $\mathbf{H}^{l+1}_\mathbf{V}$, $\mathbf{H}^{l+1}_{\boldsymbol{\phi}}$ and $\mathbf{H}^{l+1}_{\tilde{\mathbf{u}}}$ can be calculated similarly. By substituting $\delta^l$ and $\mathbf{Z}_{\mathbf{W}} ^l$ into \eqref{HW} and combining the inverse Hessian approximations with \eqref{d}, the update direction can be calculated by a two-loop recursive procedure \cite{lbfgs,lrbfgs}. The computation process is summarized as \textbf{Algorithm 1}.
	\begin{algorithm}[t]
		\caption{RBFGS algorithm with limited memory}
		\begin{algorithmic}[1]
			\Require Initial direction $\mathbf{p}^l \!=\! \operatorname{grad} g(\mathbf{X}^l)$, $m (m\!\leq\! S)$ stored medium variables $\mathbf{M}_i = (\mathbf{S}^i_{\mathbf{X}},\mathbf{Y}^i_{\mathbf{X}},\delta^i), i=1,\dots,m$. 
			\For{$i= m : -1 : 1$}
			\State $\varrho^i= \delta^i \langle \mathbf{S}^i_{\mathbf{X}} , \mathbf{p}^l\rangle$;
			\State $\mathbf{p}^l = \mathbf{p}^l - \varrho^i \mathbf{Y}_{\mathbf{X}}^i$;
			\EndFor
			\State $\mathbf{p}^l = \frac{ \langle \mathbf{S}_{\mathbf{X}}^{l-1},\mathbf{Y}_{\mathbf{X}}^{l-1}\rangle}{ \langle \mathbf{Y}_{\mathbf{X}}^{l-1},\mathbf{Y}_{\mathbf{X}}^{l-1}\rangle} \mathbf{p}^l$;
			\For{$i = 1: 1 : m$}
			\State $\beta = \delta^i \langle \mathbf{Y}_{\mathbf{X}}^i, \mathbf{p}^l \rangle$;		
			\State $\mathbf{p}^l = \mathbf{p}^l + (\varrho^i - \beta)\mathbf{S}_{\mathbf{X}}^i$;
			\EndFor
			\Ensure $\mathbf{d}^l = -\mathbf{p}^l$
		\end{algorithmic}
	\end{algorithm}
	
	With the update direction, the variable can be updated as $\mathbf{X}^{l+1} = \mathcal{R}_{\mathbf{X}^l}(\alpha^l\mathbf{d}_{\mathbf{X}}^l)$. To ensure monotonicity, we utilize the line-search update strategy, which is implemented as
	\begin{equation}	
		g(\mathbf{X}^{l+1}) \leq g(\mathbf{X}^{l}) +\sigma \gamma^n \tau_l \langle\operatorname{grad}g(\mathbf{P}^{l}), \mathbf{d}^{l} \rangle,
		\label{Amijo}
	\end{equation}
	where $\tau_l$ is a initial step size with relative large value, $\sigma$ and $\gamma \in (0,1)$ are update parameters. A proper step size can be obtained as $\alpha^l = \gamma^n\tau_l$ by increasing $n$ until \eqref{Amijo} is satisfied\cite{rmo}. The medium variables $(\mathbf{S}_\mathbf{X}^l,\mathbf{Y}_{\mathbf{X}}^l,\delta^l)$ can be obtained once $\mathbf{X}^{l}$ is updated as $\mathbf{X}^{l+1}$, and we implement cautious update to decide whether to store the medium variables for the following iterations to ensure symmetric positive definiteness of the inverse Hessian approximation \cite{rbfgs}. The cautious update condition is
	\begin{equation}
		\label{cautious}
		{\langle \mathbf{S}_\mathbf{X}^l,\mathbf{Y}_\mathbf{X}^l \rangle} \geq 10^{-4} {\langle \mathbf{S}_\mathbf{X}^l,\mathbf{S}_\mathbf{X}^l \rangle}\lVert\operatorname{grad}g(\mathbf{X}^l)\rVert.
	\end{equation}
	
	It should be noted that when implementing Algorithm 1 in the next iteration, the medium variables $(\mathbf{S}_\mathbf{X}^i,\mathbf{Y}_\mathbf{X}^i,\delta^i),\forall i\neq l$ stored in the memory should be transported to the tangent space of the current point $\mathbf{X}^{l+1}$. Due to the limited memory size, the earliest stored medium variable set should be cleared if the memory is full. The overall RBFGS for tackling \eqref{Qs} over the constructed PRMS is summarized in \textbf{Algorithm 2}.
	
	\begin{algorithm}[t]
		\caption{The RBFGS algorithm for solving \eqref{Qs}.}
		\begin{algorithmic}[1]
			\Require Initial point $\mathbf{X}^0$, empty memory $\mathbf{M}$, convergence threshold $\epsilon$, maximum iteration number $\operatorname{Maxiter}$.
			\For{$l=0: \operatorname{Maxiter}$}
			\State Obtain $\mathbf{d}_\mathbf{X}^l$ by \textbf{Algorithm 1} with $\mathbf{M}$;
			\State Obtain $\alpha^l$ and update $\mathbf{X}^{l+1}$ by \eqref{Amijo};
			\If{$\lVert\mathbf{X}^{l+1}-\mathbf{X}^{l}\rVert \leq \epsilon$}
			\State $\operatorname{Break}$;
			\EndIf
			\State Obtain $(\mathbf{S}^l_\mathbf{X},\mathbf{Y}^l_\mathbf{X},\delta^l)$ by \eqref{s} and \eqref{y};
			\State Check the number of the stored memories $m$;
			\If{\eqref{cautious} is satisfied}
			\If{$m=\mathrm{M}$}
			\State Store $\mathbf{M}_{m}=(\mathbf{S}^l_\mathbf{X},\mathbf{Y}^l_\mathbf{X},\delta^l)$;
			\For{$i=1:m-1$}
			\State $\mathbf{S}_\mathbf{X}^i = \mathcal{T}_{\mathbf{X}^{l+1}}\mathbf{S}_\mathbf{X}^{i+1}$, $\mathbf{Y}_\mathbf{X}^i = \mathcal{T}_{\mathbf{X}^{l+1}}\mathbf{Y}_\mathbf{X}^{i+1}$;
			\State $\mathbf{M}_i = (\mathbf{S}^i_\mathbf{X},\mathbf{Y}^i_\mathbf{X},\delta^{i+1})$;
			\EndFor			
			\Else
			\State Store $\mathbf{M}_{m+1}=(\mathbf{S}^l_\mathbf{X},\mathbf{Y}^l_\mathbf{X},\delta^l)$ in $\mathbf{M}$;
			\For{$i=1:m$}
			\State $\mathbf{S}_\mathbf{X}^i = \mathcal{T}_{\mathbf{X}^{l+1}}\mathbf{S}_\mathbf{X}^i$, $\mathbf{Y}_\mathbf{X}^i = \mathcal{T}_{\mathbf{X}^{l+1}}\mathbf{Y}_\mathbf{X}^i$;
			\State $\mathbf{M}_i = (\mathbf{S}^i_\mathbf{X},\mathbf{Y}^i_\mathbf{X},\delta^i)$;
			\EndFor				
			\EndIf
			\Else
			\For{$i=1:m$}
			\State $\mathbf{S}_\mathbf{X}^i = \mathcal{T}_{\mathbf{X}^{l+1}}\mathbf{S}_\mathbf{X}^i$, $\mathbf{Y}_\mathbf{X}^i = \mathcal{T}_{\mathbf{X}^{l+1}}\mathbf{Y}_\mathbf{X}^i$;
			\EndFor
			\EndIf
			\State $l=l+1$;
			\EndFor
			\Ensure $\mathbf{X} = \mathbf{X}^{l+1}$.
		\end{algorithmic}
	\end{algorithm}

	\subsection{Exact Penalty Method for Finding Feasible Solution}
	For obtaining a feasible solution of \eqref{Q2} via solving \eqref{Qs}, the values of the penalty weight $\rho$ and smoothing parameter $u$ in \eqref{Qs} are crucial. The inequality constraints are likely to be satisfied if $\rho$ is sufficiently large and $u$ is small enough. However, too large $\rho$ and small $u$ lead to slow convergence and numerical difficulties. In contrast, the obtained solution of $\eqref{Qs}$ can be far from feasible if $\rho$ is too small. A practical approach is the exact penalty method, where a feasible solution is found by iteratively solving \eqref{Qs} via Algorithm 2 and updating $\rho$ and $u$ based on whether the inequalities are satisfied, until the constraints are satisfied with convergence \cite{rep}. Specifically, after obtaining the solution of \eqref{Qs}, $\rho$ is increased as $\rho=\theta_\rho \rho$ with $\theta_\rho >1$ if \eqref{ineq} is not satisfied. The smoothing parameter is decreased iteratively as $u=\max\{u_{\min},\theta_u u\}$ with $\theta_u \in (0,1)$ to increase the precision of the linear-quadratic function, where $u_{\min}$ is the lower bound. Based on the exact penalty method and Algorithm 2, the PRMO for solving \eqref{Q3} is summarized as \textbf{Algorithm 3}.
	
	The penalty weight $\rho$ and smoothing parameter $u$ can be fixed once the proper values are found. The line-search strategy \eqref{Amijo} guarantees the monotonicity of Algorithm 2 \cite{conv}, that is if $\mathbf{X}^{l}$ is bounded over the $\mathcal{M}$, there exists $c > 0$  such that
	\begin{equation}	
		g(\mathbf{X}^{l+1}) - g(\mathbf{X}^{l}) \leq   c \langle\operatorname{grad}g(\mathbf{X}^{l}), \mathbf{d}_\mathbf{X}^{l} \rangle,
	\end{equation}
	then we can conclude that $g(\mathbf{X}^{l+1}) \leq g(\mathbf{X}^{l})$ and $g(\mathbf{X})$ is non-increasing, and the power must be positive. Thus, with proper $\rho$ and $u$, Algorithm 3 is guaranteed to converge.
	
	The starting point is important for improving the efficiency of the proposed PRMO for power minimization. The initial positions can be obtained by the circle packing scheme to guarantee that the positions are sufficiently separated \cite{MA4}. The IRS reflection coefficients and receive filters can be randomly initialized, then we initialize beamforming $\mathbf{W}$ via the SDR technique with one time Gaussian randomization to obtain an initial point that is relatively close to the optimal point \cite{sdr}. The SDR method is illustrated in Appendix B.
	
	\begin{algorithm}[t]
		\caption{PRMO method for the power minimization}
		\begin{algorithmic}[1]
			\Require Initial point $\mathbf{X}^1$, initial parameters $\rho^1$, $u^1$, $\epsilon^1$, $\theta_\rho>1$, $\theta_u\in(0,1)$, $\theta_\epsilon\in(0,1)$, lower bound $u_{\min}, \epsilon_{min}$, convergence threshold $\tau$.
			\Repeat
			\State \parbox[t]{\dimexpr\linewidth-\algorithmicindent}{Obtain $\mathbf{X}^{l+1}$ by solving \eqref{Qs} via \textbf{Algorithm 2} with warm-start at $\mathbf{X}^l$, fixed penalty weight $\rho^l$, smoothing parameter $u^l$, and convergence threshold $\epsilon^l$;} 
			\If{$h_i(\mathbf{X}^{l+1})>0, \exists i \in \mathcal{I}$}
			\State $\rho^{l+1}=\theta_\rho \rho^l$;
			\Else
			\State $\rho^{l+1}=\rho^l$;
			\EndIf
			\State $u^{l+1} = \max\{\theta_u u^l,u_{\min}\}$, $\epsilon^{l+1} = \max\{\theta_\epsilon \epsilon^l,\epsilon_{\min}\}$;			
			\State $l=l+1$;
			\Until{$\lVert \mathbf{X}^{l+1}-\mathbf{X}^l\rVert < \tau$ and $h_i(\mathbf{X}^{l+1})\leq0, \forall i \in \mathcal{I}$ and $u^{l+1}\leq u_{\min}$ and $\epsilon^l\leq \epsilon_{min}$.}
			\Ensure $\mathbf{X}^{l+1}=[\mathbf{W}^{l+1},\mathbf{V}^{l+1},\boldsymbol{\phi}^{l+1},\tilde{\mathbf{u}}^{l+1}]$, $\mathbf{W} = \mathbf{W}^{l+1}$,  $\mathbf{V} = \mathbf{V}^{l+1}$, $\boldsymbol{\phi} = \boldsymbol{\phi}^{l+1}$,  $\mathbf{u}=\boldsymbol{p}(\tilde{\mathbf{u}}^{l+1})$.
		\end{algorithmic}
	\end{algorithm}

	\subsection{Comparison Between Element-Wise Control and Array-Wise Control for the MIRS}
	Note that when the MIRS adopts the element-wise control scheme, this corresponds to the special case of $a=1$, which means the positions of all the IRS elements are optimized by the PRMO. Although the element-wise control provides the highest flexibility and
	theoretically optimal performance, its computational complexity becomes
	prohibitive when the number of movable reflecting elements $N$ is large. Specifically, for the minimum distance constraint \eqref{Ct}, $\frac{N(N-1)}{2}$ inequalities and the corresponding gradients are required to be computed in each step of Algorithm 2. Although adopting the linear quadratic function can avoid the evaluation of negative inequality constraints and the corresponding gradients, the number of inequalities still grows quadratically with $N$, which leads to substantial computational overhead, especially when the number of elements becomes large. To reduce the computational overhead while exploiting the advantages of MIRS, the array-wise control scheme can be adopted by increasing $a$. We can find that compared with element-wise control, the numbers of the position variables and the corresponding inequality constraints are reduced, which lead to less computation required. Specifically, the dimensionality of the variable is reduced by a factor of $1/a^2$, while the number of minimum ME distance constraints decreases from $\frac{N(N-1)}{2}$ to $\frac{N(N-a^2)}{2a^4}$. This reduction in problem size effectively reduces the computational resources required, enabling a better trade-off between system performance and computational overhead, particularly when a large number of reflecting elements is employed. Moreover, the solution obtained by array-wise control can be indirectly applied to the element-wise MIRS optimization, for example, as an initialization for further fine-tuning of the positions.

	\subsection{Computational Complexity Analysis}
	We first focus on the element-wise control. The complexity of the proposed PRMO method mainly lies in the calculation of the Euclidean gradients and the implementation of Algorithm 2 in the inner iterations. The complexity of Algorithm 2 is analyzed in Table I.
	\begin{table}[htbp]
		\caption{Computational Complexity Analysis of \textbf{Algorithm 2}}
		\centering
		\begin{tabular}{|c|c|} \hline
			Term & Complexity Order  \\ \hline
			$\nabla_{\mathbf{W}^*}h_i(\mathbf{X}),\forall i\in\mathcal{K}_c$&  $\mathcal{O}\left(M(K_c+K_t)^2\right)$\\ \hline
			$\nabla_{\boldsymbol{\phi}*}h_i(\mathbf{X}),\forall i\in\mathcal{K}_c$&  $\mathcal{O}\left(N^2M+M(K_c+K_t)^2\right)$\\ \hline
			$\nabla_{\tilde{\mathbf{t}}}h_i(\mathbf{X}),\forall i\in\mathcal{K}_c$&  $\mathcal{O}\left(N^2M(K_c+K_t)+M(K_c+K_t)^2\right)$\\ \hline
			$\nabla_{\mathbf{W}^*}h_i(\mathbf{X}),\forall i\in\mathcal{K}_t$&  $\mathcal{O}(M^2(Kc+Kt)^2)$\\ \hline
			$\nabla_{\mathbf{V}^*}h_i(\mathbf{X}),\forall i\in\mathcal{K}_t$&  $\mathcal{O}(M^2(Kc+Kt)^2)$\\ \hline
			$\nabla_{\boldsymbol{\phi}^*}h_i(\mathbf{X}),\forall i\in\mathcal{K}_t$&  $\mathcal{O}(M^3K_t+N^2M^2K_t+K_t^2M^2)$\\ \hline
			$\nabla_{\tilde{\mathbf{t}}}h_i(\mathbf{X}),\forall i\in\mathcal{K}_t$&  $\mathcal{O}(K_tM^3+N^2K_tM)$\\ \hline
			$\nabla_{\tilde{\mathbf{t}}}h_i(\mathbf{X}),\forall i\in\mathcal{D}$&  $\mathcal{O}(N)$\\ \hline
			\textbf{Algorithm 1}  &  $\mathcal{O}\bigl(S(K_t^2M+MK_cK_t+N)\bigr)$ \\ \hline
			Total  & \makecell[c]{$\mathcal{O}\bigl(M^3K_t+N^3+K_c^4M+K_t^2M^3$\\$+K_t^2N^2M\!+\!S(M^2K_t+K_cK_tM\!+\!N)\bigr)$} \\ \hline
		\end{tabular}
	\end{table}
	
	Besides, the complexity of the SDR method for initializing $\mathbf{W}$ is $\mathcal{O}(M^{3.5}(K_c+K_t)^{3.5})$. Based on the above, the computational complexity of the PRMO summarized in Algorithm 3 is estimated as $\mathcal{O}\bigl(M^{3.5}(K_c+K_t)^{3.5}+I_1I_2(M^3K_t+N^3+K_c^3M+K_t^3M^2+S(M^2K_t+K_cK_tM+N))\bigr)$, where $I_1$ is the number of inner iterations in Algorithm 2 and $I_2$ is the number of inner iteration for implementing Algorithm 2 in Algorithm 3. 
	
	Typically, the number of IRS elements $N$ is much larger than other parameters. By implementing the array-wise control, the dimension of position variable is reduced by a factor of $\frac{1}{a^2}$, then the computation overhead associated with the position optimization, which is primarily determined by $N$, can be significantly reduced, thereby improving computational efficiency.
	
	\section{Simulation Results}
	Simulation results are provided in this section to prove the effectiveness of the proposed MIRS-aided ISAC systems. Unless otherwise noted, the simulation parameters are as follows. Consider the Cartesian coordinate with the unit of 1m, the BS and MIRS are located at $(0,0,0)$ and $(30,10,5)$, respectively. We assume that $M=12$, $N=36$, $A=8\lambda$, $K_t=2$, $K_c=2$, $L=6$, $T=1024$, $\Gamma_k=8\ \text{bps/Hz}, \forall k\in\mathcal{K}_c$, $\chi_k=12\ \text{dB}, \forall k \in\mathcal{K}_t$, $\sigma_{c,k}^2=-120$ dBm, and $\sigma_{S}^2=-110$ dBm. The CU locations are randomly chosen within a square region with $x\sim[50,55]$ and $y\sim[0,5]$ in the $x$-$y$ plane. The channel path angles $\phi_{t,l}$, $\theta_{r,l},\phi_{r,l},\forall l$, and the target AoD $\theta_{I,k},\phi_{I,k},\forall k\in\mathcal{K}_t$ are randomly generated from uniform distribution over $[0,\pi]$. The carrier frequency is set as 2 GHz, and the corresponding free space path loss is $L(d) = -38-20 \log (d)$ dB, where $d$ is the channel distance. Considering LoS component in the BS-MIRS channel, The channel response $\boldsymbol{\Sigma}_G$ is generated by $\sigma_{G,1}\sim\mathcal{CN}(0,\mu_G\kappa/(\kappa+1))$ and $\sigma_{G,l}\sim\mathcal{CN}(0,\mu_G/(L-1)(\kappa+1)),\forall l=2,\dots,L$, where $\kappa=6$ dB is the ratio of the average power
	of LoS component to that of non-LoS component. For the MIRS-CU channels, we set $\sigma_{k,l}\sim\mathcal{CN}(0,\mu_k/L)$. $\mu_G$ and $\mu_k$ are the path losses of the BS-IRS channel and the IRS-CU channel for the $k$-th CU, respectively. The maximum inner iteration number of Algorithm 2 is $100$.
	\begin{figure}[t]
		\centering{\includegraphics[width=0.8\columnwidth]{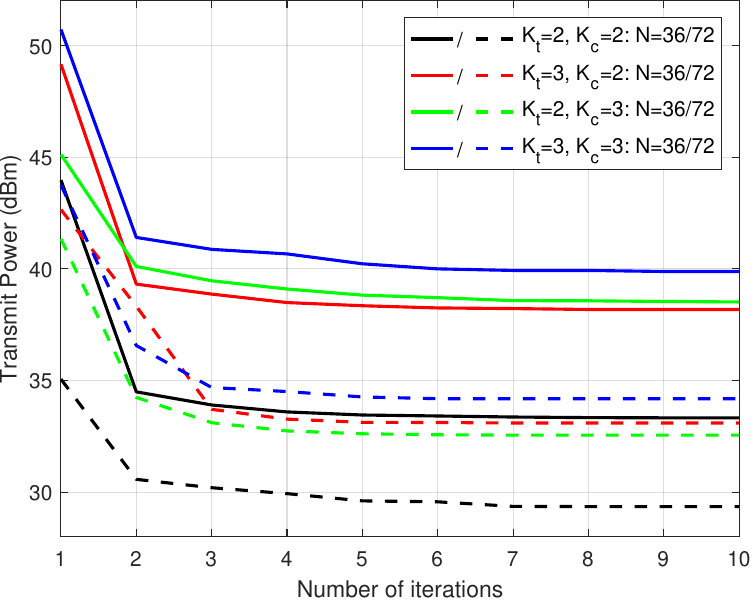}}
		\caption{Convergence of Algorithm 3 with different numbers of CU and targets.}
	\end{figure}
	\begin{figure}[t]
		\centering{\includegraphics[width=0.8\columnwidth]{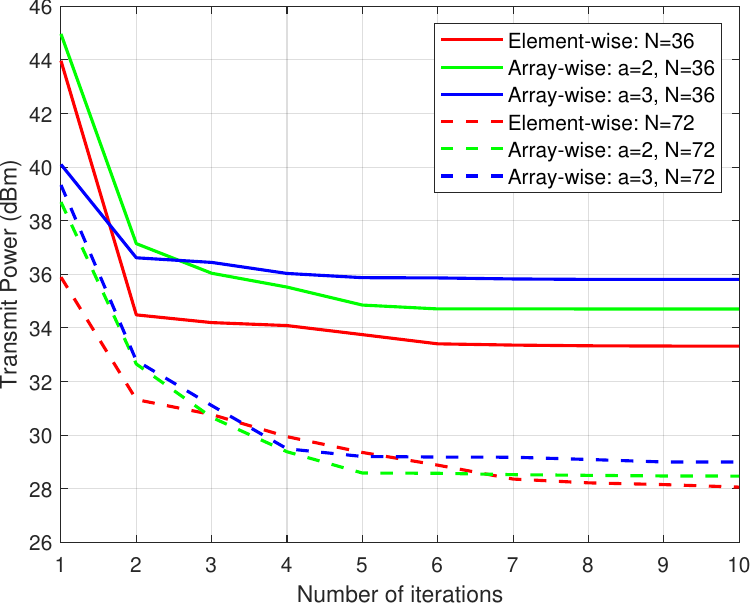}}
		\caption{Convergence of Algorithm 3 under element-wise control and array-wise control.}
	\end{figure}
	\subsection{Convergence Performance}
	
	In this section, we evaluate the convergence performance of Algorithm 3 with different system parameters. Figure 3 illustrates the convergence performance of Algorithm 3 with varying numbers of CUs, targets, and MIRS elements. It can be observed that with same number of MIRS elements, a larger number of targets or CUs leads to an increased number of iterations required for convergence. In particular, when $N=72$ and $K_t=K_c=2$, Algorithm 3 converges within 5 iterations, whereas when $K_t=K_c=3$, 7 iterations are needed to reach convergence. Moreover, as the number of CUs or targets increases, the optimized transmit power also increases, since more communication and sensing threshold constraints need to be satisfied. Besides, by increasing the number of MIRS elements, the optimized transmit power decreases due to the increased DoFs that can be exploited. 
	
	Figure 4 evaluates the convergence behavior of the proposed MIRS under element-wise and array-wise control for different $N$, while Table II shows the corresponding execution time and achieved transmit power. We can observe that although element-wise control achieves lower transmit power, it requires more iterations due to the large number of inequality constraints, which increases the computational overhead of the Riemannian gradient calculations in Algorithm 2. In contrast, the proposed array-wise control reduces both the number of constraints and the dimension of position variables, enabling faster convergence with significantly lower computational cost and only minor performance loss, especially when the number of reflecting elements is large. Specifically, when $K_c = K_t = 2$ and $N = 36$, adopting the array-wise control scheme with $a = 2$ reduces the execution time by $66.1\%$ while increasing the transmit power by only $4.2\%$ compared with the element-wise control scheme. When $a = 3$, the execution time is reduced by $72.9\%$ with a transmit power increase of $7.5\%$. Furthermore, when the number of reflecting elements increases to $N = 72$, the array-wise control schemes with $a = 2$ and $a = 3$ reduce the execution time by $85.7\%$ and $88.3\%$, respectively, while the corresponding transmit power increases are only $1.5\%$ and $3.3\%$. These results demonstrate that array-wise control can achieve a favorable performance--overhead trade-off with moderate array sizes, especially when the number of reflecting elements becomes large.

	\begin{table}[t]
		\caption{Comparison of Optimized Transmit Power /Execution Time with Convergence}
		\centering
		\begin{tabular}{|c|c|c|} \hline
			Scheme & \!\!$K_t=2,K_c=2$\!\!&\!\!$K_t=3,K_c=3$\!\!\\ \hline
			MIRS-EC: $N=36$ & \!\!\!33.32 dBm / 58.44 s\!\!\!	&\!\!39.87 dBm / 84.17 s\!\! \\ \hline
			\!\!MIRS-AC: $a\!\!=\!\!2, N\!\!=\!\!36$\!\!&\!\!34.71 dBm / 19.79 s\!\!& \!\!41.19 dBm / 47.63 s \!\!\\ \hline
			\!\!MIRS-AC: $a\!\!=\!\!3, N\!\!=\!\!36$\!\!&\!\!35.81 dBm / 15.80 s\!\!& \!\!42.66 dBm / 31.32 s \!\!\\ \hline
			MIRS-EC: $N=72$ & \!\!28.06 dBm / 212.53 s\!\!	& 34.18 dBm / 278.49 s\\ \hline
			\!\!MIRS-AC: $a\!\!=\!\!2, N\!\!=\!\!72$\!\!& \!\!28.47 dBm / 30.43 s\!\! & \!\!35.75 dBm / 39.20 s\!\! \\ \hline
			\!\!MIRS-AC: $a\!\!=\!\!3, N\!\!=\!\!72$\!\!& \!\!28.99 dBm / 24.86 s\!\! & \!\!36.97 dBm / 28.12 s\!\! \\ \hline
		\end{tabular}
	\end{table}
	\subsection{Performance Comparison}
	In this section, we provide the performance comparison to evaluate the effectiveness of the proposed MIRS-aided ISAC system. The proposed MIRS with element-wise control and array-wise control are denoted as ``\textbf{MIRS-EC}" and ``\textbf{MIRS-AC}", respectively. Besides, we consider the following schemes for comparisons: 
	\begin{itemize}
		\item ``\textbf{BF Only}": Only the beamforming $\mathbf{W}$ and receive filter $\mathbf{V}$ at BS are optimized via the PRMO. FPA-IRS with random reflection coefficients is adopted. Based on Table I, the computational complexity of the scheme is  $\mathcal{O}\bigl(M^{3.5}(K_c+K_t)^{3.5}+I_1I_2(M^3K_t+K_c^3M+K_t^3M^2+S(M^2K_t+K_cK_tM))\bigr)$
		\item ``\textbf{FPA-IRS}": $\mathbf{W}$, $\mathbf{V}$, and $\boldsymbol{\phi}$ of the FPA-IRS are jointly optimized via the PRMO. Based on Table I, the computational complexity of the scheme is $\mathcal{O}\bigl(M^{3.5}(K_c+K_t)^{3.5}+I_1I_2(M^3K_t+N^2M^2K_t^2+K_c^3M+K_t^3M^2+S(M^2K_t+K_cK_tM+N))\bigr)$
		\item ``\textbf{Sep}": The variables are separately optimized via existing optimization techniques to obtain a feasible solution. Specifically, the element positions and $\boldsymbol{\phi}$ are obtained to maximize the sum channel gain via gradient descent \cite{MA3}, $\mathbf{V}$ is obtained by the Rayleigh quotation \cite{IRS3}, and $\mathbf{W}$ is optimized via the SDR method with $500$ times Gaussian randomization \cite{sdr}. The complexity of the scheme is $\mathcal{O}\bigl(M^{3.5}(K_c+K_t)^{3.5}+N^2(K_c+K_t+M)\bigr)$.
	\end{itemize}
	\begin{figure}[t]
		\centering{\includegraphics[width=0.8\columnwidth]{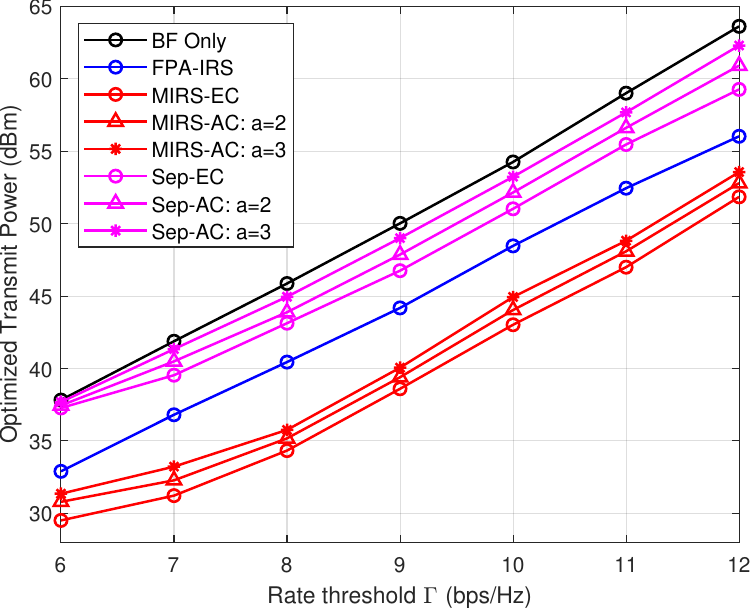}}
		\caption{Transmit power versus communication rate threshold $\Gamma$ .}
	\end{figure}
	\begin{figure}[t]
		\centering{\includegraphics[width=0.8\columnwidth]{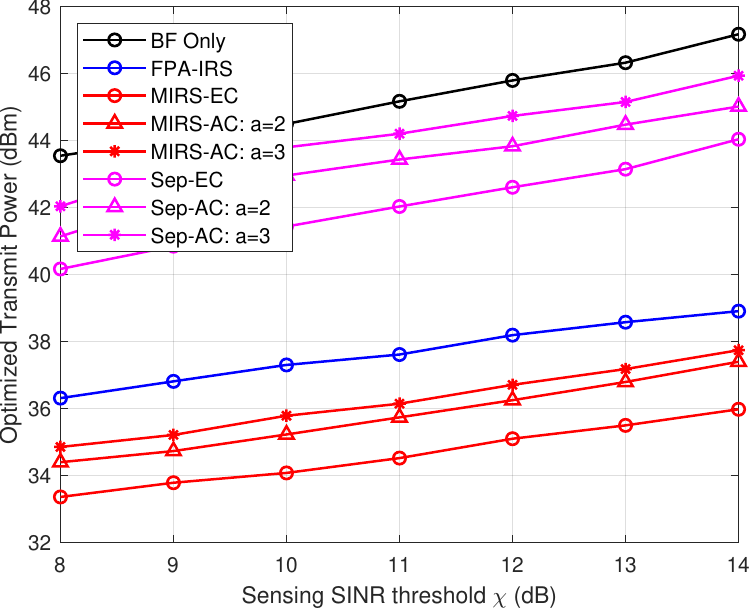}}
		\caption{Transmit power versus sensing SINR threshold $\chi$.}
	\end{figure}
	
	Figures 5 and 6 illustrate the impact of communication rate threshold $\Gamma$ and sensing SINR threshold $\chi$ on the optimized transmit power. As shown in Fig. 5, a higher $\Gamma$ results in an increased optimized transmit power. By employing traditional IRS with optimized reflection coefficients, the required transmit power can be reduced compared with optimization based solely on the active beamforming at BS. Furthermore, exploiting the position optimization of the movable elements, the proposed MIRS achieves a lower transmit power. The results also indicate that the proposed PRMO that jointly optimizes the variables outperforms the method that optimizes the variables separately via existing methods. For different control schemes of the MIRS, the element-wise control scheme achieves the lowest optimized transmit power, and the array-wise control schemes achieve relatively higher transmit power, while still outperforming the conventional FPA-IRS scheme. Similarly, Fig. 6 shows that an increase in the sensing SINR threshold also leads to higher transmit power requirements. Moreover, compared with the FPA-IRS scheme, the proposed MIRS further reduces the required transmit power by optimizing the element positions with element-wise control and array-wise control schemes.
	\begin{figure}[t]
		\centering{\includegraphics[width=0.8\columnwidth]{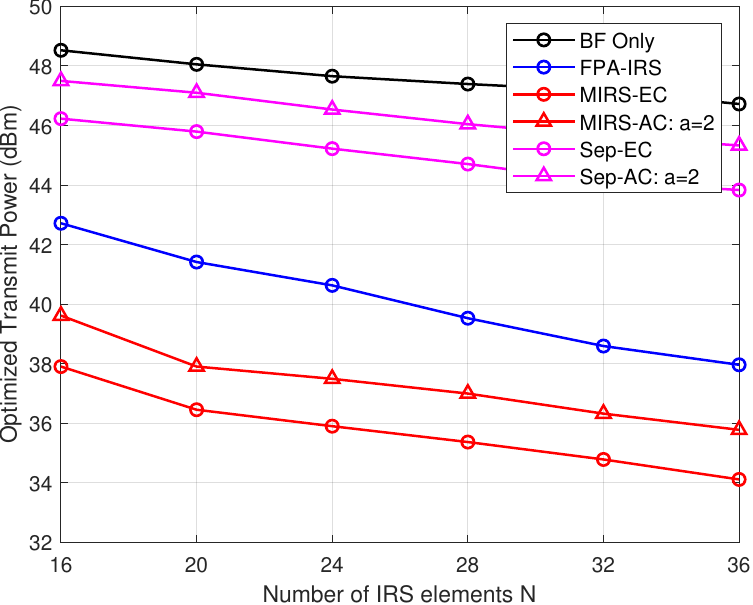}}
		\caption{Transmit power versus number of reflecting elements.}
	\end{figure}
	
	\begin{figure}[t]
		\centering{\includegraphics[width=0.8\columnwidth]{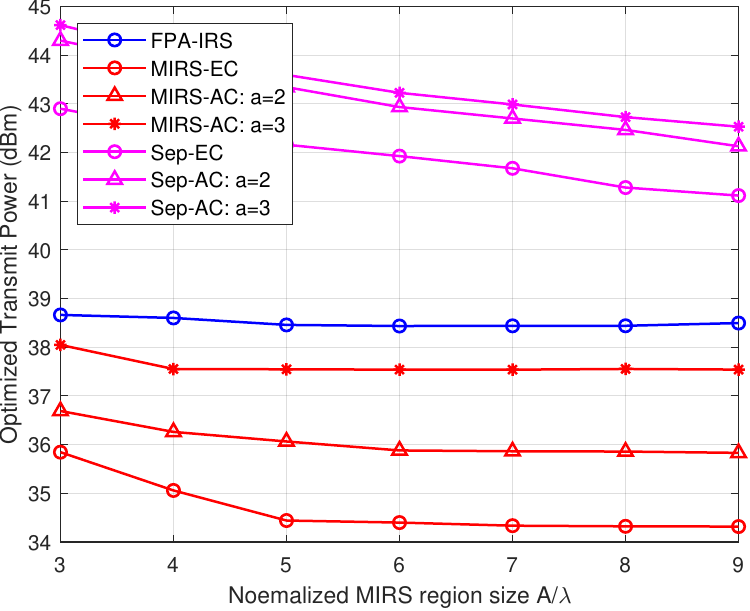}}
		\caption{Transmit power versus normalized IRS region size $A/\lambda$.}
	\end{figure}
	
	The impact of the numbers of reflecting elements of conventional FPA-IRS and the proposed MIRS on ISAC performance is demonstrated in Fig. 7. As $N$ increases, the optimized transmit power of all schemes decreases due to higher array gain and enhanced channel control. Moreover, MIRS achieves lower transmit power compared with the conventional FPA-IRS by exploiting additional spatial DoFs. 
	
	It is worth noting that the minimum distance constraint imposed on elements limits their mobility when the number of elements becomes large, thereby reducing the performance gain over FPA-IRS. This issue can be mitigated by enlarging the MIRS region size. Figure 8 presents the influence of the MIRS region size on system performance. It is shown that increasing the region size leads to a reduction in optimized transmit power. When the size of the MIRS becomes sufficiently large, each element can be positioned to a location with favorable channel condition while satisfying the minimum distance constraint, and the transmit power gradually converges.
	
	\begin{figure}[t]
		\centering{\includegraphics[width=0.8\columnwidth]{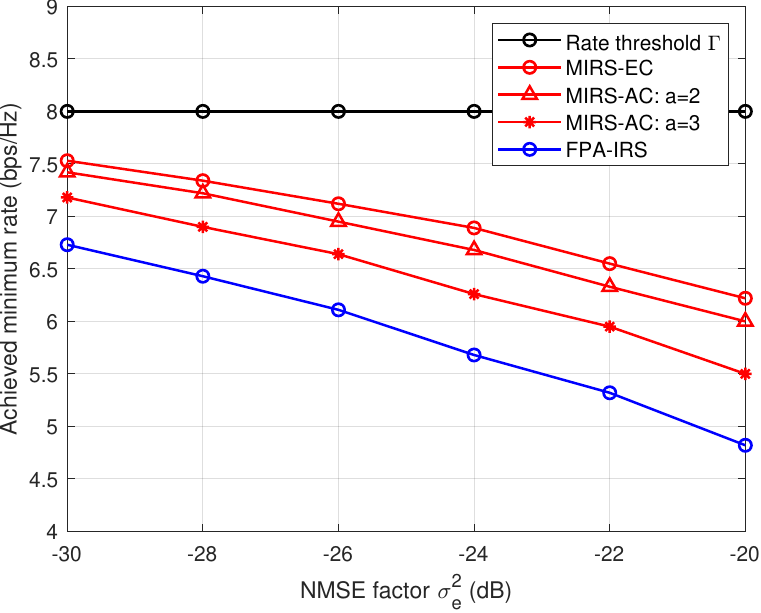}}
		\caption{Achieved minimum communication rate versus NMSE $\sigma_e^2$.}
	\end{figure}
	
	\subsection{Impact of Imperfect CSI}
	The aforementioned simulation results and discussions assume perfect CSI to evaluate the theoretical performance of the proposed MIRS-aided ISAC systems. However, as noted in Section II-B, random multipath interference arising from uncorrelated scattering is often present in wireless environments, which poses challenges for accurate estimation using existing CE methods. Furthermore, target mobility may introduce errors in the target channel model. Consequently, this section evaluates the robustness and performance of the proposed PRMO algorithm under imperfect CSI conditions.
	\begin{figure}[t]
		\centering{\includegraphics[width=0.8\columnwidth]{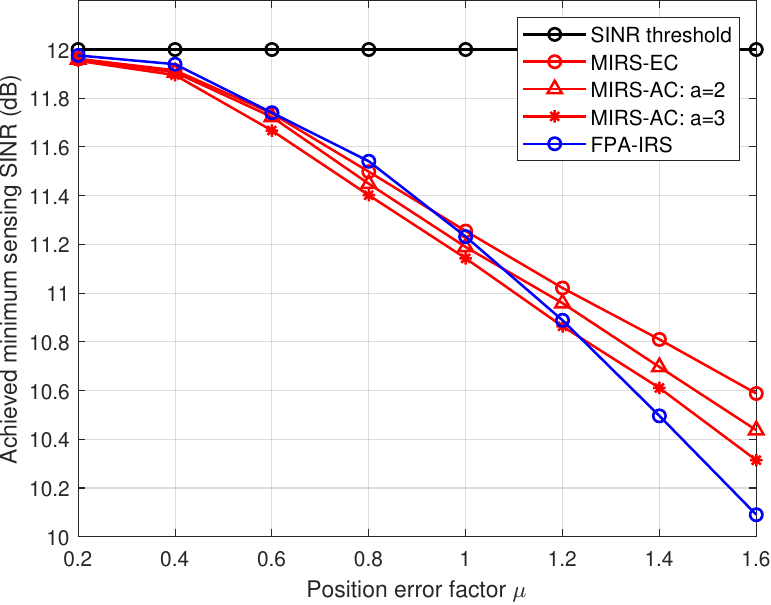}}
		\caption{Achieved minimum sensing SINR versus position error factor $\mu$.}
	\end{figure}
	
	\subsubsection{Impact of random multipath component}
	In this section, we evaluate the impact of uncorrelated multipath interference on MIRS-aided ISAC systems. Since the BS and MIRS are stationary, the BS-MIRS channel is assumed to be perfectly known. In contrast, estimation errors exist in the MIRS-CU channels due to the random distribution of CUs. Let $\bar{\mathbf{f}}_k$ and $\tilde{\mathbf{f}}_k, \forall k\in\mathcal{K}_c$ denote the estimated MIRS-CU channels for optimization and the true MIRS-CU channels, respectively. The estimated channel is modeled as
	\begin{align}
		\bar{\mathbf{f}}_k = \tilde{\mathbf{f}}_k + \mathbf{e},\forall k\in\mathcal{K}_c,
	\end{align}
	where $\mathbf{e}\sim\mathcal{CN}(0,\sigma_e^2\frac{\lVert\tilde{\mathbf{f}}_k\rVert_2^2}{N})$ represents the estimation error induced by uncorrelated scattering, and $\sigma_e^2$ is the normalized mean square error (NMSE). For the power minimization problem, the solution obtained under imperfect CSI may result in communication rates that fall below the predefined threshold $\Gamma$. Figure 9 illustrates the minimum achieved communication rates versus various NMSE values for different schemes. It is observed that the communication rates degrade as the NMSE increases. Notably, even with lower optimized transmit power, the proposed MIRS under both element-wise and array-wise control schemes exhibits less performance loss compared to the conventional FPA-IRS. This demonstrates that the proposed MIRS architecture is more robust and effective than conventional FPA-IRS for the communications of the ISAC.
	
	\subsubsection{Impact of imperfect target position information}
	As discussed in Section II-C, the target positions estimated in the previous sensing stage can be utilized for channel modeling in the current stage. However, these positions may be inaccurate due to target mobility. This section evaluates the effectiveness of the proposed MIRS scheme under imperfect target position information. Denote the position of the $k$-th target for obtaining its AoDs with respect to the BS and MIRS and the corresponding channels as $\bar{\mathbf{p}}_k\in\mathbb{R}^3,\forall k\in\mathcal{K}_t$. The true position of the $k$-th target is $\tilde{\mathbf{p}}_k=[\tilde{p}_{k,1},\tilde{p}_{k,2},\tilde{p}_{k,3}]^T\in\mathbb{R}^3$. The target position  $\bar{\mathbf{p}}_k=[\bar{p}_{k,1},\bar{p}_{k,2},\bar{p}_{k,3}]^T\in\mathbb{R}^3$ which is utilized for optimization is modeled such that each element follows a uniform distribution  $\bar{p}_{k,n}\sim\mathcal{U}(\tilde{p}_{k,n}-\mu/2,\tilde{p}_{k,n}+\mu/2),\forall n \in \{1,2,3\}$, where $\mu$ represents the error factor. Figure 10 demonstrates the impact of the error factor $\mu$ on the minimum sensing SINR for the targets under different schemes. It is observed that while imperfect position information leads to SINRs below the predefined threshold, the performance degradation is limited for all schemes with relatively small error factor. It indicates that the sensing is effective for the targets which are static or moving slowly. Notably, at higher error factors, the proposed MIRS with both element-wise and array-wise control exhibits smaller performance losses compared to the FPA-IRS. This demonstrates the enhanced robustness of the MIRS architecture against errors in the CSI of the targets.

	\section{Conclusion}
	In this paper, we have proposed a novel MIRS-aided ISAC system. Compared with conventional FPA-IRS, MIRS enhances the ISAC performance by controlling the element position and exploiting additional spatial degrees of freedom. Considering the trade-off between system performance and computational overhead, two modes, including element-wise control and array-wise control, are designed for the MIRS. A PRMO method was then developed to jointly optimize the positions with other system parameters including beamforming, receive filters, and IRS reflection coefficients. Simulation results have demonstrated the advantages of the MIRS compared with conventional FPA-IRS, as well as the effectiveness of the PRMO method compared with existing optimization algorithms. Theoretical performance of the MIRS-aided ISAC system was mainly focused on in this paper, and more practical problems, such as more effective and low-complexity algorithm for the real-time and rapid control of element positions, could be studied in future works.

	\appendices
	\section{Detailed Partial Derivatives}
	Exploiting basic matrix differentiation \cite{CM}, for $h_i(\mathbf{X}),\forall i\in\mathcal{K}_c$, we have
	\begin{flalign}	
		\frac{\partial r_i}{\partial \mathbf{W}^*} = \left[\frac{\partial r_i}{\partial \mathbf{w}_1^*},\dots,\frac{\partial r_i}{\partial \mathbf{w}_{K_c+K_t}^*}\right],
	\end{flalign}
	\begin{flalign}	
		\frac{\partial r_i}{\partial \boldsymbol{\phi}^*}=\frac{\partial r_i}{\partial \gamma_{C,i}}\frac{\partial \gamma_{C,i}}{\partial \boldsymbol{\phi}^*} = \frac{2}{1+\gamma_{C,i}}\cdot\frac{\partial \gamma_{C,i}}{\partial \boldsymbol{\phi}^*},
	\end{flalign}
	and
	\begin{flalign}	
		\frac{\partial r_i}{\partial \tilde{\mathbf{t}}} =\mathcal{D}_{\mathbf{t}}^Tr_i\odot A\mathbf{sig}(\tilde{\mathbf{t}})\odot (1-\mathbf{sig}(\tilde{\mathbf{t}})),
	\end{flalign}
	where
	\begin{flalign}	
		\frac{\partial r_i}{\partial \mathbf{w}^*_i} = \frac{\partial r_i}{\partial \gamma_i}\frac{\partial \gamma_i}{\partial \mathbf{w}^*_i} = \frac{2}{1+\gamma_i}\cdot\frac{\mathbf{h}_{C,i}^H\mathbf{w}_i\mathbf{h}_{C,i}}{{\sum_{j\neq k}\lvert\mathbf{h}^H_{C,i}\mathbf{w}_j\rvert^2 + \sigma_n^2}},
	\end{flalign}
	\begin{flalign}	
		\frac{\partial r_i}{\partial \mathbf{w}^*_j} = \frac{2}{1+\gamma_i}\cdot\frac{-\lvert\mathbf{h}^H_{C,i}\mathbf{w}_i\rvert^2\mathbf{h}_{C,i}^H\mathbf{w}_j\mathbf{h}_{C,i}}{\left({\sum_{j\neq i}\lvert\mathbf{h}^H_{C,i}\mathbf{w}_j\rvert^2 + \sigma_n^2}\right)^2}, \forall j \neq i,
	\end{flalign}
	\begin{flalign}	
		\frac{\partial \gamma_{C,i}}{\partial \boldsymbol{\phi}^*}\!=\!\left(\mathcal{D}_{\boldsymbol{\phi}^*}\mathbf{h}_{C,i}\right)^T\frac{\partial \gamma_{C,i}}{\partial \mathbf{h}_{C,i}},\ \mathcal{D}_{\boldsymbol{\phi}^*}\mathbf{h}_{C,i} \!=\! \mathbf{G}^T\operatorname{diag}(\mathbf{f}^H_{i}) ,
	\end{flalign}
	\begin{align}	
		&\frac{\partial \gamma_{C,i}}{\partial \mathbf{h}_{C,i}}\!=\! \frac{\mathbf{w}_i^H\mathbf{h}_{C,i}\mathbf{w}_i}{{\sum_{j\neq i}\lvert\mathbf{h}^H_{C,i}\mathbf{w}_j\rvert^2 + \sigma_n^2}}\nonumber\\&\qquad\qquad\qquad\qquad- \sum_{j\neq i}\frac{\lvert\mathbf{h}^H_{C,i}\mathbf{w}_i\rvert^2\mathbf{w}_j^H\mathbf{h}_{C,i}\mathbf{w}_j}{\left({\sum_{j\neq i}\lvert\mathbf{h}^H_{C,i}\mathbf{w}_j\rvert^2 + \sigma_n^2}\right)^2},
	\end{align}
	and
	\begin{align}
		\mathcal{D}_{\mathbf{t}}r_i\!\!=\!\! \frac{2}{1\!+\!\gamma_{C,i}}\!\!\left(\!\!\operatorname{vec}^T\!\!\left(\frac{\partial \gamma_{C,i}}{\partial \mathbf{G}^*}\!\!\right)\!\!\mathcal{D}_{\mathbf{t}}\!\mathbf{G}^*\!\!\!+\!\operatorname{vec}^T\!\!\left(\!\!\frac{\partial \gamma_{C,i}}{\partial \mathbf{f}_i^*}\right)\!\!\mathcal{D}_{\mathbf{t}}\mathbf{f}_i^*\!\!\right)\!\!,
	\end{align}
	where
	\begin{align}	
		\frac{\partial \gamma_{C,i}}{\partial \mathbf{G}^*}=& \frac{\mathbf{h}_{C,i}^H\mathbf{w}_i\operatorname{diag}(\mathbf{f}^H_i)\boldsymbol{\phi}\mathbf{w}_i^H}{{\sum_{j\neq i}\lvert\mathbf{h}^H_{C,i}\mathbf{w}_j\rvert^2 + \sigma_n^2}}-\nonumber\\ &\qquad\quad\sum_{j\neq i}\frac{\lvert\mathbf{h}_{C,i}^H\mathbf{w}_i\rvert^2\mathbf{h}_{C,i}^H\mathbf{w}_j\operatorname{diag}(\mathbf{f}^H_i)\boldsymbol{\phi}\mathbf{w}_j^H}{\left({\sum_{j\neq i}\lvert\mathbf{h}^H_{C,i}\mathbf{w}_j\rvert^2 + \sigma_n^2}\right)^2},
	\end{align}
	\begin{align}	
		\frac{\partial \gamma_{C,i}}{\partial \mathbf{f}_i^*}=& \frac{\mathbf{w}_i^H\mathbf{h}_{C,i}\operatorname{diag}(\boldsymbol{\phi}^H)\mathbf{G}\mathbf{w}_i^H}{{\sum_{j\neq i}\lvert\mathbf{h}^H_{C,i}\mathbf{w}_j\rvert^2 + \sigma_n^2}}-\nonumber\\ &\qquad\quad\sum_{j\neq i}\frac{\lvert\mathbf{h}_{C,i}^H\mathbf{w}_i\rvert^2\mathbf{w}^H_j\mathbf{h}_i\operatorname{diag}(\boldsymbol{\phi}^H)\mathbf{G}\mathbf{w}_j^H}{\left({\sum_{j\neq i}\lvert\mathbf{h}^H_{C,i}\mathbf{w}_j\rvert^2 + \sigma_n^2}\right)^2}.
	\end{align}
	For the term $\mathcal{D}_\mathbf{t}\mathbf{G}^*$, the $(N(m-1)+n)$-th row of which is$\frac{\partial \mathbf{G}^*[n,m]}{\partial \mathbf{t}}=\left[\mathbf{0}^T,\dots,\mathbf{0}^T,\left(\frac{\partial \mathbf{G}^*[n,m]}{\partial \mathbf{t}_n}\right)^T,\mathbf{0}^T,\dots,\mathbf{0}^T\right]$,
	where $\frac{\partial \mathbf{G}^*[n,m]}{\partial \mathbf{t}_n}=\sum_{l=1}^{L} \jmath\frac{2\pi}{\lambda}\sigma_{G,l}^*e^{\jmath\frac{2\pi}{\lambda}(\boldsymbol{\rho}_{r,l}^T\mathbf{t}_n-{\rho}_{t,l}{b}_m)}\boldsymbol{\rho}_{r,l}$. 
	Then $\mathcal{D}_\mathbf{t}\mathbf{f}_i^*$ can be obtained as the same way.
	
	For $h_i(\mathbf{X}),i\in\mathcal{K}_t$, we define $n_i=K_t\sigma^2_n\mathbf{v}_i^H\mathbf{v}_i$ and $\tilde{\mathbf{V}_i}\in\mathbb{C}^{M\times(K_c+K_t)}$ with $\mathbf{v}_i=\operatorname{vec}(\tilde{\mathbf{V}_i})$, then we have
	\begin{align}
		\partial\gamma_{S,i}/{\partial \mathbf{v}_i^*}\!=\!\frac{\mathbf{A}_i\!\mathbf{v}_i\!+\!\mathbf{A}_i^H\!\mathbf{v}_i}{\mathbf{v}_i^H\mathbf{B}_i\mathbf{v}_i}\!\!-\!\!\frac{\mathbf{v}_i^H\!\mathbf{A}_i\!\mathbf{v}_i\mathbf{B}_i\mathbf{v}_i\!+\!\mathbf{v}_i^H\!\mathbf{A}^H_i\!\mathbf{v}_i\mathbf{B}^H_i\mathbf{v}_i}{(\mathbf{v}_i^H\mathbf{B}_i\mathbf{v}_i)^2},
	\end{align}
	\begin{align}
		&\operatorname{vec}\left(\partial\gamma_{S,i}/{\partial\mathbf{W}^*}\right)=\partial\gamma_{S,i}/{\partial\mathbf{w}^*}\nonumber\\&\ \ = \!\frac{\mathbf{C}_i\!\mathbf{w}\!+\!\mathbf{C}_i^H\!\mathbf{w}}{\mathbf{w}^H\mathbf{D}_i\mathbf{w}+n_i}\!-\!\frac{\mathbf{w}^H\!\mathbf{C}_i\!\mathbf{w}\mathbf{D}_i\mathbf{w}\!+\!\mathbf{w}^H\!\mathbf{C}^H_i\!\mathbf{w}\mathbf{D}^H_i\mathbf{w}}{(\mathbf{w}^H\mathbf{D}_i\mathbf{w}+n_i)^2},
	\end{align}
	\begin{align}
		(\partial\gamma_{S,i}/\partial{\boldsymbol{\phi}^*})^T=&\mathcal{D}_{\mathbf{H}_{S,i}}\gamma_{S,i}\mathcal{D}_{\boldsymbol{\phi}^*}\mathbf{H}_{S,k}\nonumber\\&+\!\!\!\sum_{j\in\mathcal{K}_t,j\neq i}\!\!\!\left(\frac{\partial\gamma_{S,i}}{\partial\operatorname{vec}(\mathbf{H}_{S,j})}\right)^T\mathcal{D}_{\boldsymbol{\phi}^*}\mathbf{H}_{S,j},
	\end{align}
	and $\nabla_{\tilde{\mathbf{t}}}\gamma_{S,i} =\nabla_{\mathbf{t}}\gamma_{S,i}\odot A\mathbf{sig}(\tilde{\mathbf{t}})\odot (1-\mathbf{sig}(\tilde{\mathbf{t}}))$, where 
	\begin{align}
		&\frac{\partial\gamma_{S,i}}{\partial{\mathbf{t}}}\!\!=\!\!\mathcal{D}_{\mathbf{H}_{S,i}}\!\!\gamma_{S,i}\mathcal{D}_{\mathbf{h}^*_{S,i}}\!\!\mathbf{H}_{S,i}\bigl(\!\mathcal{D}_{\mathbf{G}^T}\!\mathbf{h}^*_{S,i}\mathcal{D}_{\mathbf{t}}\mathbf{G}^T\!\!+\!\!\mathcal{D}_{\mathbf{h}^*_{r,i}}\!\!\mathbf{h}^*_{S,i}\mathcal{D}_{\mathbf{t}}\mathbf{h}^*_{r,i}\bigr)\nonumber\\&+\!\!\!\!\!\!\sum_{j\in \mathcal{K}_t,j\neq i}\!\!\!\!\!\!\!\mathcal{D}_{\mathbf{H}_{S,j}}\!\!\gamma_{S,i}\mathcal{D}_{\mathbf{h}^*_{S,j}}\!\!\mathbf{H}_{S,j}\bigl(\!\mathcal{D}_{\mathbf{G}^T}\!\mathbf{h}^*_{S,j}\mathcal{D}_{\mathbf{t}}\mathbf{G}^T\!\!+\!\!\mathcal{D}_{\mathbf{h}^*_{r,j}}\!\!\mathbf{h}^*_{S,j}\mathcal{D}_{\mathbf{t}}\mathbf{h}^*_{r,j}\bigr),
	\end{align}
	and terms in the expressions are obtained as $\mathbf{A}_i=T\operatorname{vec}(\mathbf{H}_{S,i}\mathbf{W})\operatorname{vec}(\mathbf{H}_{S,i}\mathbf{W})^H$,
	$\mathbf{B}_i=\sum_{j\in\mathcal{K}_t,j\neq i}\!T\operatorname{vec}(\mathbf{H}_{S,j}\mathbf{W})\operatorname{vec}(\mathbf{H}_{S,j}\mathbf{W})^H+\sigma^2_S$, $
	\mathbf{C}_i=T\operatorname{vec}(\mathbf{H}_{S,i}\tilde{\mathbf{V}}_i)\operatorname{vec}(\mathbf{H}_{S,i}\tilde{\mathbf{V}}_i)$, $
	\mathbf{D}_i=\sum_{j\in\mathcal{K}_t,j\neq i}T\operatorname{vec}(\mathbf{H}_{S,j}\tilde{\mathbf{V}}_i)\operatorname{vec}(\mathbf{H}_{S,j}\tilde{\mathbf{V}}_i)$, $
	\mathcal{D}_{\mathbf{H}_{S,i}}\gamma_{S,i}\!=\!\frac{\partial\gamma_{S,i}}{\partial\operatorname{vec}(\mathbf{H}_{S,i})}\!=\!2T\mathbf{v}_i^T\operatorname{vec}^*(\mathbf{H}_{S,i}\tilde{\mathbf{V}}_i)\operatorname{vec}(\tilde{\mathbf{V}}^*_{i}{\mathbf{W}}^T)/n_i$, $
	\mathcal{D}_{\mathbf{H}_{S,j}}\gamma_{S,i}\!=\!\frac{\partial\gamma_{S,i}}{\partial\operatorname{vec}(\mathbf{H}_{S,j})}\!=\!-2T\mathbf{v}_i^T\operatorname{vec}^*(\mathbf{H}_{S,j}{\mathbf{W}}) \operatorname{vec}^T(\mathbf{H}_{S,i})$ $ \operatorname{vec}(\tilde{\mathbf{V}}^*_{i}{\mathbf{W}}^T)\mathbf{v}_i^T\operatorname{vec}^*(\mathbf{H}_{S,i}{\mathbf{W}})\operatorname{vec}(\tilde{\mathbf{V}}^*_{i}{\mathbf{W}}^T)\mathbf{v}_i^T/n_i$, $
	\mathcal{D}_{\boldsymbol{\phi}^*}\mathbf{H}_{S,i}=(\mathbf{h}^*_{S,i}\otimes\mathbf{I}_M+\mathbf{I}_M\otimes\mathbf{h}^*_{S,i})\mathbf{G}^T\operatorname{diag}(\mathbf{h}^H_{r,i})$, $
	n_i\!=\!T\sum_{j\neq i}\alpha_j^2\lvert\mathbf{v}_i^H\left(\mathbf{I}_{K_c+K_t}\otimes\mathbf{H}_{S,j}\right)\mathbf{w}\rvert^2+K_t\sigma_S^2\mathbf{v}_i^H\mathbf{v}_i$, $
	\frac{\partial \mathbf{h}_{r,i}}{\partial\mathbf{t}_n}=-\jmath\frac{2\pi}{\lambda}\sqrt{\alpha_{r,i}}e^{-\jmath\frac{2\pi}{\lambda}\boldsymbol{\rho}^T_{I,i}\mathbf{t}_n}\boldsymbol{\rho}_{I,i}$, 
	$\mathcal{D}_{\mathbf{h}^*_{S,i}}\mathbf{H}_{S,i}=\mathbf{h}^*_{S,i}\otimes\mathbf{I}_M+\mathbf{I}_M\otimes\mathbf{h}^*_{S,i}$, $\mathcal{D}_{\mathbf{G}^T}\mathbf{h}^*_{S,i}=\mathbf{h}^H_{r,i}\operatorname{diag}(\boldsymbol{\phi})\otimes\mathbf{I}_M$, $\mathcal{D}_{\mathbf{h}^*_{r,i}}\mathbf{h}^*_{S,i}=\mathbf{G}^T\operatorname{diag}(\boldsymbol{\phi})$.
	
	\section{SDR for Initializing the Beamforming Matrix}
	Define $\mathbf{P}_k\!\!=\!\!\mathbf{A}_k-g_k\mathbf{B}_k,\forall k \in\mathcal{K}_c$, $\mathbf{Q}_k\!\!=\!\!\mathbf{C}_k - \chi\mathbf{B}_k, \forall k \in\mathcal{K}_t$, and $c_k=\sigma_S^2\mathbf{v}^H_k\mathbf{v}_k,\forall k\in\mathcal{K}_t$, where $\mathbf{A}_k=\boldsymbol{\Xi}_k^H\mathbf{h}_{C,k}\mathbf{h}_{C,k}^H\boldsymbol{\Xi}_k$, $\mathbf{B}_k=\sum_{j\neq k}\boldsymbol{\Xi}_j^H\mathbf{h}_{C,k}\mathbf{h}_{C,k}^H\boldsymbol{\Xi}_j$, $g_k=e^{\Gamma_k} -1$, $\boldsymbol{\Xi}_k=[\mathbf{0}_{M\times (k-1)M}\ \mathbf{I}_M\ \mathbf{0}_{M\times (M(K_c+K_t)-kM)}]$, $\mathbf{C}_k=T(\mathbf{I}_{K_c+K_t}\otimes\mathbf{H}^H_{S,k})\mathbf{v}_k\mathbf{v}_k^H(\mathbf{I}_{K_c+K_t}\otimes\mathbf{H}_{S,k})$, and $\mathbf{D}_k=\sum_{j\neq k}T(\mathbf{I}_{K_c+K_t}\otimes\mathbf{H}^H_{S,j})\mathbf{v}_k\mathbf{v}_k^H(\mathbf{I}_{K_c+K_t}\otimes\mathbf{H}_{S,j})$, when optimizing $\mathbf{W}$ while keep other variables fixed, define $\bar{\mathbf{W}}=\mathbf{w}\mathbf{w}^H$, the subproblem with respect to $\mathbf{w}$ based on \eqref{Q} is
	\begin{subequations}
		\label{sdp}
		\begin{IEEEeqnarray}{r,l}			
			$$\underset{\bar{\mathbf{W}}}{\min}$$&{\ \operatorname{Tr}(\bar{\mathbf{W}})}\\
			$$\operatorname{s.t.}$$ 
			&\ \operatorname{Tr}(\mathbf{P}_k\bar{\mathbf{W}})\leq g_k\times \sigma_{c,k}^2, \forall k \in\mathcal{K}_c,\\
			&\ \operatorname{Tr}(\mathbf{Q}_k\bar{\mathbf{W}})\leq \chi_k\times c_k, \forall k\in\mathcal{K}_t,\\
			&\ \operatorname{rank}(\bar{\mathbf{W}})=1 \label{rank1}.
		\end{IEEEeqnarray}
	\end{subequations}
	Then, by relaxing the rank 1 constraint \eqref{rank1}, \eqref{sdp} can be solved via standard convex optimization solvers. Then, a feasible $\mathbf{w}$ could be obtained via the eigen-decomposition method or the Gaussian randomization technique \cite{sdr}.

	\bibliographystyle{IEEEtran}
	\bibliography{Refs}{}

\end{document}